\documentclass{article}
\usepackage[margin=1in]{geometry}
\usepackage{graphicx} 
\usepackage{todonotes}
\usepackage{amsmath}
\usepackage{amssymb}
\usepackage{amsthm}
\usepackage{hyperref}

\usepackage{empheq}
\usepackage{booktabs}
\usepackage{multirow}
\usepackage{colortbl}
\usepackage{booktabs}
\usepackage{tikz}
\usetikzlibrary{calc}
\usepackage{threeparttable}
\usepackage[ruled,vlined]{algorithm2e}
\usepackage{subcaption}
\usepackage{authblk}

\usepackage[numbers,square, sort]{natbib}
\title{Controllable Generation of Implied Volatility Surfaces with Variational Autoencoders}

\author[1]{Jing Wang}
\author[1,2]{Shuaiqiang Liu \footnote{The views expressed in this paper are  personal views of the author and do not necessarily reflect the views or policies of his current or past employers.}}
\author[1]{Cornelis Vuik}

\affil[1]{Numerical Analysis, Delft University of Technology, Delft, the Netherlands}
\affil[2]{ING Bank, Amsterdam, the Netherlands}

\date{\today}

\begin{document}

\maketitle
\begin{abstract}
This paper presents a deep generative modeling framework for controllably synthesizing implied volatility surfaces (IVSs) using a variational autoencoder (VAE). Unlike conventional data-driven models, our approach provides explicit control over meaningful shape features (e.g., volatility level, slope, curvature, term-structure) to generate IVSs with desired characteristics.
In our framework, financially interpretable shape features are disentangled from residual latent factors. The target features are embedded into the VAE architecture as controllable latent variables, while the residual latent variables capture additional structure to preserve IVS shape diversity. To enable this control, IVS feature values are quantified via regression at an anchor point and incorporated into the decoder to steer generation.
Numerical experiments demonstrate that the generative model enables rapid generation of realistic IVSs with desired features rather than arbitrary patterns,  and achieves high accuracy across both single- and multi-feature control settings. For market validity, an optional post-generation latent-space repair algorithm adjusts only the residual latent variables to remove occasional violations of static no-arbitrage conditions without altering the specified features.
Compared with black-box generators, the framework combines interpretability, controllability, and flexibility for synthetic IVS generation and scenario design.

\end{abstract}

\tableofcontents

\section{Introduction}

The generation of realistic and diverse market scenarios plays a crucial role in quantitative finance, underpinning applications such as risk management, regulatory compliance, and trading strategy evaluation. See \citep{SOA2016ESG} for further  applications with generated economic scenarios.  These scenarios, whether derived from historical data or constructed hypothetically, serve as  representations of potential market conditions. They enable financial institutions to evaluate a broad range of possible outcomes under different assumptions. A critical requirement in this context is controllability: the ability to guide the scenario generation process so that the resulting  data exhibit desired and financially meaningful features. Without such control, generated scenarios may display arbitrary patterns of limited practical use.  For example, the importance of controllability is particularly evident in regulatory contexts, where scenario-based stress testing has become a standard tool for resilience assessment. Article 177 of the Capital Requirements Regulation (CRR) \citep{EUReg2013CRR} mandates the use of “severe but plausible” recession scenarios when evaluating capital adequacy. Those specified features require a controllable generator of  market scenarios.

In option markets, scenarios are most naturally represented through implied volatility surfaces (IVSs). Black-Scholes implied volatility, obtained by inverting the Black–Scholes pricing formula to match an observed option quote, encodes the market’s assessment of future uncertainty under the risk-neutral measure.  An IVS describes how implied volatilities vary across strike prices and maturities, and  provides a standardized and comprehensive presentation of the option market. Importantly, its shape characteristics, such as the overall volatility level and skewness, are economically interpretable and provide insight into prevailing market conditions. These features are particularly suitable as control variables to generate synthetic IVSs with desired features. This paper focuses on the controllable generation of synthetic IVSs.

Beyond the methodological interest, a controllable generator for IVSs can serve several purposes in quantitative finance. It enables targeted scenario construction, for instance,  by generating surfaces with elevated volatility levels to simulate stress periods \citep{malz2001financial}. It can also provide a complementary tool for risk measurement: synthetic IVSs can be used to compute risk metrics (e.g. Value at Risk) when historical data are insufficient, or be used to impute missing option quotes in illiquid markets.  Additionally, controllable generators can be used for downstream tasks, for instance,  the data-driven market simulator for deep hedging \citep{buehler2019deep},  where large amounts of synthetic but realistic IVS data are needed.

\subsection{Literature review}
Methods for generating IVSs can be broadly categorized into \emph{parametric} and \emph{data-driven} approaches. Parametric methods typically employ stochastic differential equations to model the dynamics of asset prices and volatility processes, or closed-form functional specifications designed to ensure arbitrage-free conditions. Representative examples include stochastic volatility models such as SABR \citep{hagan2002managing}, Heston \citep{heston1993closed} and rough volatility models \citep{Bayer02062016},  parametric forms such as Stochastic Volatility Inspired (SVI) model \cite{gatheral2014arbitrage}, and non-parametric  forms such as local volatility models \citep{dupire1994pricing}. These approaches are interpretable but rely on strong assumptions about market dynamics, which can limit the diversity of admissible IVS shapes. In contrast, data-driven methods learn IVS features directly from observed  data. They impose fewer a priori constraints and offer greater flexibility. This flexibility makes them particularly promising for generating synthetic IVSs for broader market scenarios.  See the review \citep{horvath2025generative} for generating financial market data (not limited to IVS) with deep learning-based techniques.

Generative modeling provides a general framework for data-driven scenario generation. In this setting, each IVS can be represented as a high-dimensional data object, assumed to be drawn from some unknown distribution. With only a finite collection of observed surfaces available, the objective is to approximate this distribution by learning a mapping from a simple latent distribution (e.g., Gaussian) to the true distribution of IVSs. A parameterized generator function carries out this mapping, so that drawing a random sample from the latent distribution and applying the generator produces new synthetic IVSs. The generator is trained so that the distribution of these synthetic surfaces approximates the distribution of the observed data.

Classical generative models such as Gaussian mixture models \citep{kienitz2024gaussian} and PCA-based probability models \citep{Cont04032023} are simple and interpretable but limited in their ability to deal with large-scale IVSs. For example,  the paper \citep{Cont04032023} stated that the selected  principal components do not cover all of the variance in the corresponding data.  However, deep generative models, such as Variational Autoencoders (VAEs) \citep{kingma2014semi}, Generative Adversarial Networks (GANs) \citep{goodfellow2014generative}, and diffusion models \citep{ho2020denoising}, leverage deep neural networks to capture complex patterns. Diffusion models often produce samples of very high quality, but they require computationally expensive, multi-step generation procedures. By contrast, VAEs and GANs can generate new samples in a single step.

For IVS modeling, GAN-based approaches (e.g., volatility-GANs \citep{Vuletic03072024} and the arbitrage-minimal GAN of \citep{na2023computingvolatilitysurfacesusing}) have demonstrated the feasibility of generating realistic volatility surfaces. In comparison, VAEs offer several distinct advantages over GANs in this setting. First, VAEs include an explicit encoder network. This allows efficient inference of latent variables from observed data and helps in analyzing and interpreting surface structures. Second, VAEs operate in a compact lower-dimension latent space. This makes it possible to create a direct link between financial interpretations and learned representations, which improves interpretability. Third, VAEs typically exhibit more stable training behavior compared to GANs.

A serial of  studies have explored VAE-based IVS generation. \citep{bergeron2021variational} were the first to model IVSs using a purely data-driven VAE approach, contrasting grid-based and pointwise architectures on FX options data. However, their method did not enforce arbitrage constraints or interpret latent dimensions. \citep{ning2023arbitrage} introduced a hybrid method that enforces arbitrage by combining VAEs with stochastic differential equation (SDE) models, but the generated surfaces remain dependent on the chosen SDE specification (e.g., regime-switching or Levy additive processes), limiting their purely data-driven nature. \citep{kunsagi2023deep} investigated latent space representations but found that interpretations were dataset-specific and often entangled across dimensions. More recently, \citep{gopal2024filling} focused on distributional imputation of IVSs using VAEs, finding that residual network architectures and learnable VAEs outperform standard $\beta$-VAE formulations \citep{higgins2017beta}, though their emphasis was on imputation rather than controlled generation.

\subsection{Our work}

Despite growing interest in generative modeling of IVSs, existing approaches generally lack a mechanism for controllable generation, that is,  the ability to produce surfaces with specified, financially meaningful features. Addressing this gap is one of the main motivations for the controllable VAE developed in this paper.  
In our formulation, IVSs are encoded into latent space that is split into two components: (i) controllable variables corresponding to target surface attributes, and (ii) residual latent variables capturing other variation. This design enables the generation of surfaces that satisfy targeted specifications while preserving diversity.

When considering target features of IVSs, a natural question arises: which features are suitable for control, and how should they be quantified? Although an IVS is inherently high-dimensional, empirical studies have shown that its variation can be captured by a small number of stylized factors, for instance, skew and smile effects and the term structure of volatility in \citep{cont2002dynamics}, long memory in volatility in \citep{Bayer02062016}, and local concavity in \citep{franccois2022venturing}. These features are both intuitive and financially interpretable, for example, the skew steepened during financial crises \citep{constantinides2015supply}, which makes them suitable candidates for controllable variables.
In this paper,   we focus on four shape characteristics as proof of concept: at-the-money volatility level, slope and curvature along strike prices, term structure along time to maturity. Note that our method is not restricted to the above features.  To quantify these features, we select an anchor point at the at-the-money strike as time to maturity approaches zero. The related limiting behavior has been studied extensively, for example,  different option pricing models exhibiting distinct signatures at this point \citep{carr2003type, ait2021implied}. Conversely, given an observed IVS, \citep{ait2021implied} proposed regression-based methods to estimate shape features at this point and relate them to suitable option pricing models when calibration to market data. Inspired by this idea, we employ a bivariate regression approach to extract the selected features at the anchor point and incorporate them into the controllable generation process.

In addition, any generated surface must respect no-arbitrage constraints, in particular the absence of butterfly and calendar spread arbitrage. While our controllable VAE enforces feature-level structure, arbitrage violations can still arise when latent variables are sampled in the tails of the normal distribution. To address this issue, we develop a repair algorithm that adjusts problematic surfaces to minimize arbitrage violations while preserving their overall shape.

Overall, this paper bridges scenario design and generative modeling. By combining interpretability, controllability, and arbitrage-free conditions, our approach provides a data-driven modelling framework for generating IVSs that are both economically meaningful and practically useful.

The remainder of this paper is organized as follows. In \autoref{sec:IVS}, we provide a brief overview regarding implied volatility surfaces. In \autoref{sec:ShapeFeatures} we describe the anchor point and the quantification of IVS shape characteristics. \autoref{sec:CVAE} describes the proposed controllable generative modelling framework based on a variational auto-encoder and quantitative IVS features. \autoref{sec:Numerical} presents  numerical experiments. \autoref{sec:conclusion} concludes.

\section{Implied Volatility Surfaces (IVS)}\label{sec:IVS}

We review the construction of implied volatility surfaces (IVS), beginning with the 
Black--Scholes framework and the definition of implied volatility. We then describe how 
IVSs are derived from market data and conclude with the no-arbitrage conditions that any 
valid surface must satisfy. These elements provide the financial foundation for the 
generative modeling framework developed in later sections.

\subsection{Black--Scholes model and implied volatility}

In the Black--Scholes (BS) model, the underlying asset price follows
\begin{subequations}\label{eq:BS}
\begin{align}
\mathrm{d}S(t) &= rS(t)\,\mathrm{d}t + \sigma S(t)\,\mathrm{d}W^{\mathbb{Q}}(t),
\quad S(t_0) = S_0 > 0,
\end{align}
\end{subequations}
where $S(t)$ is the underlying asset price with initial value $S_0$, $r$ the 
risk--free rate, $\sigma$ the constant volatility parameter, $W^{\mathbb{Q}}(t)$ a 
Brownian motion under the risk--neutral measure, and $K$ the strike price. A European call option has  payoff $V_c(T, S) := \max(S(T)-K, 0)$ at maturity time $T$, and the corresponding option  price at time $t$ before maturity reads
\begin{subequations}\label{eq:analyticalBScall}
\begin{align}
V_c(t, S) &= S \mathcal{N}(d_1) - K e^{-r\tau} \mathcal{N}(d_2),\\
d_1 &= \frac{\log(S/K) + (r + 0.5\sigma^2)\tau}{\sigma\sqrt{\tau}}, 
\quad d_2 = d_1 - \sigma\sqrt{\tau},
\end{align}
\end{subequations}
where $\tau := T-t$ is the time to maturity, and $\mathcal{N}(\cdot)$ the standard 
normal cumulative distribution function. By the put-call parity, the price of vanilla European put options  can be obtained analytically. 

For each observed market option price $V_{\text{market}}$, 
 there exists a unique volatility value $\sigma$ that makes the BS formula 
match it. This parameter is called the \emph{implied volatility}. Implied 
volatility provides a standardized way to compare options across strikes and maturities 
and serves as the basic element of an implied volatility surface.

\subsection{Construction of IVS}

The construction of an IVS in practice involves three steps:
\begin{enumerate}
    \item Collect market option prices $V_{\text{market}}$ together with strikes $K$ 
    and time to maturities $\tau$;
    \item Compute the implied volatility $\sigma$ by inverting the BS option pricing model, 
    $\sigma = BS^{-1}(V_{\text{market}}; S_0, K, \tau, r)$;
    \item Interpolate or smooth the resulting data points of implied volatilities to obtain a surface 
    $\sigma(m,\tau)$, spanning on log--moneyness $m=\log(K/S_0)$  
    and time to maturity $\tau$.
\end{enumerate}

Empirical studies show that IVSs display systematic patterns such as smiles, skews, 
and term-structure effects. These can be effectively summarized by a small number of 
shape features. We defer the formal definition and quantification of these features to 
Section~\ref{sec:ShapeFeatures}, where they will play an important role in our controllable generative modeling of IVSs.

\subsection{Arbitrage-free conditions}
\label{sec:arbitrage_free_conditions}

Any valid option price surface must satisfy no-arbitrage conditions to prevent 
inconsistencies across strikes and maturities.  
Referring to the paper~\citep{gatheral2014arbitrage}, we can express these conditions 
in terms of the total implied variance,
\begin{equation}\label{eq:TTV}
    w(m, \tau) := \sigma^2(m, \tau) \cdot \tau.
\end{equation}
While asymptotic no-arbitrage conditions at limiting strikes or maturities are necessary in general~\citep{gatheral2014arbitrage},   IVSs are typically studied on a finite domain of $(m,\tau)$ in practice, which is the focus of this work. On such domains, the following two conditions are most relevant.

\paragraph{Calendar spread arbitrage.}
A volatility surface is free of calendar spread arbitrage if the total variance is non-decreasing in maturity:
\begin{equation}
    \frac{\partial w(m, \tau)}{\partial \tau} \geq 0, 
    \quad \forall m \in \mathbb{R}, \; \tau > 0.
\end{equation}

\paragraph{Butterfly arbitrage.} \label{par:butterfly}
A volatility surface is said to be \textit{free of butterfly arbitrage} at maturity $\tau$ if, for the fixed maturity $\tau > 0$, the function $m \mapsto w(m; \tau)$ corresponds to a non-negative risk-neutral density.

To make this precise, consider the Black-Scholes formula for the price of a European call option with log-moneyness $m \in \mathbb{R}$ and total implied variance $w(m; \tau)$ at fixed maturity $\tau > 0$:
\begin{equation}
C_{\text{BS}}(m, w(m; \tau)) = S \left( \mathcal{N}(d_+(m; \tau)) - e^{m} \mathcal{N}(d_-(m; \tau)) \right),
\end{equation}
where $\mathcal{N}(\cdot)$ denotes the standard Gaussian cumulative distribution function, and
\begin{equation}
d_\pm(m; \tau) := \frac{-m}{\sqrt{w(m; \tau)}} \pm \frac{\sqrt{w(m; \tau)}}{2}.
\end{equation}

For a fixed $\tau > 0$, define the function $f : \mathbb{R} \to \mathbb{R}$ by
\begin{equation}\label{eq:butterfly}
f(m; \tau) := \left( 1 - \frac{m \, \partial_m w(m; \tau)}{2 w(m; \tau)} \right)^2
- \frac{(\partial_m w(m; \tau))^2}{4} \left( \frac{1}{w(m; \tau)} + \frac{1}{4} \right)
+ \frac{1}{2} \, \partial_{mm} w(m; \tau).
\end{equation}
The volatility surface is free of butterfly arbitrage at $\tau$ if and only if $f(m; \tau) \geq 0$ for all $m \in \mathbb{R}$, and
\begin{equation}
\lim_{m \to +\infty} d_+(m; \tau) = -\infty.
\end{equation}

These conditions guarantee  an IVS is free of static arbitrage within the finite domain considered in this work. In Section~\ref{sec:exp3}, we will use them to check and, 
if needed, adjust generated IVSs in order to avoid arbitrage violations.

\section{IVS shape features}
\label{sec:ShapeFeatures}
A central element of our approach is the identification of shape features of IVSs that are both financially interpretable and suitable for controllable generation. In this section, we identify these features  and describe how they can be quantified. Empirical studies on stock  and foreign exchange options, as documented in \citep{cont2002dynamics,franccois2022venturing,CHALAMANDARIS2011623}  among others,  have shown that  IVSs can be approximately represented by a small number of dominant stylized factors,  for instance,  the volatility level, slope, curvature, and term structure.  More recent research has identified additional features, for instance, local concave short-term implied volatility curves caused by event risks such as earnings announcements \citep{alexiou2025pricing}. While many features can be considered, we focus here on four canonical ones as a proof of concept.

\subsection{Shape features to control}
In this paper, we select four features as controllable variables: (i) the volatility level, (ii) slope across strikes, (iii) curvature across strikes, and (iv) the slope of the term structure with respect to maturity.

\paragraph{Volatility level.}  
The level factor reflects the overall height of the surface and is typically measured by ATM volatility. It corresponds to the market's risk-neutral expectation of future variance,  and provides the foundation for volatility indices such as the VIX \citep{gatheral2006volatility} based on S\&P 500 index options. 

\paragraph{Slope across strikes.}  
The slope describes the first-order variation of implied volatility with respect to log-moneyness. In equity index options, the slope is generally negative, implying higher implied volatilities for out-of-the-money puts relative to calls. This feature reflects the leverage effect, the persistent demand for downside protection, and is consistent with negative risk-neutral skewness \citep{bollen2004does}. Importantly, the skew steepened during the financial crises \citep{constantinides2015supply},

\paragraph{Curvature across strikes.}  
The curvature factor captures the convexity of the volatility smile, i.e., the second-order variation of implied volatility across strikes. Elevated curvature indicates that extreme strikes are priced with relatively higher volatilities, which is consistent with fat-tailed risk-neutral distributions and compensation for jump or kurtosis risk \citep{lee2002moment}.

\paragraph{Term-structure slope.}  
The term-structure factor measures the variation of implied volatility across maturities. An upward-sloping structure (contango) is typically observed in calm markets, indicating that short-term variance is below its long-run mean. A downward-sloping structure  appears in periods of stress, when short-term volatility rises sharply above long-term expectations. This behavior is closely linked to the forward variance curve and the term structure of variance swap rates \citep{egloff2010term}.

These shape features are both intuitive and financially interpretable,  which makes them suitable candidates for controllable variables. 
Although our proposed controllable generation method is demonstrated on these features, the framework is not restricted to them.

\subsection{Anchor Point for Feature Quantification}

To obtain quantitative measures of these features, we introduce an anchor point, i.e., a limiting location on the IVS used to define and extract shape characteristics. Following the literature \citep{ait2021implied, alos2021malliavin}, we choose the origin of an IVS, corresponding to the limit $(\tau \to 0^+, m = 0)$.

This choice is motivated by two considerations. First, the asymptotic behavior of IV curves near expiry has been extensively analyzed and provides theoretical discrimination between different asset price models. For instance, \citep{carr2003type} has shown that the convergence rate of option prices as $\tau \to 0$ reveals whether the underlying dynamics are continuous, discontinuous, or mixed. The recent book \citep{ait2021implied} rigorously studies   the ATM short-time  implied volatility by means of Malliavin Calculus  and compared different limiting behavior of various pricing models (e.g.,  local, stochastic, and stochastic-local volatilities,  and  rough volatilities models). Similarly, \citep{AITSAHALIA2021364} demonstrates how local shapes of an observed IVS at this point constrain the class of stochastic volatility models consistent with market data.   Second, given the information in this region, IVS  can be  reconstructed to some extent via a  combination of these factors,   without using any stochastic models \citep{franccois2022venturing}.

Other approaches such as Principal Component Analysis (PCA) have also been used to extract shape characteristics  from observed IVSs. For example, \cite{Cont04032023} decompose daily S\&P 500 IVSs as

\begin{equation*}
    \sigma_t(m, \tau) \approx \bar{\sigma}(m, \tau) \cdot \exp\left( \sum_{i=1}^k Y_t^i f_i(m, \tau) \right),
\end{equation*}
where $( m = \log(K/S_0) $ denotes log-moneyness and $ \tau = T - t $ is the time to maturity. The function $ \bar{\sigma}(m, \tau) $ represents the time-averaged IVS, while $ f_i(m, \tau) $ are the $i$-th principal components, typically interpreted as eigenfunctions of the empirical covariance operator. The time-varying coefficients $ Y_t^i $ control the contribution of each shape component. Specifically, $ Y_t^1 $ affects the overall level, $ Y_t^2 $ governs the skewness, $ Y_t^3 $ controls the term-structure, and $ Y_t^4 $ influences the curvature around the at-the-money region. However, these features of resulting IVS  are not quantitatively measured, which makes it difficult to establish a precise  connection  between a  specific feature  of  and the corresponding latent factor, when generating  desired IVS features for target market scenario.

As the paper \citep{Cont04032023} stated that the selected  principal components do not cover all of the variance in the corresponding data. The PCA is a dimensionality reduction method rather than a generative model. Although PCA-based models can be extended into generative frameworks by combining them with stochastic processes (e.g., Ornstein–Uhlenbeck dynamics on the PCA scores), the resulting surfaces  lack explicit and quantitative control over specific shape characteristics.

This limitation also motivates the nonlinear, (explicit) feature-controlled generative approach developed in this paper. While various methods for extracting such features exist, the novelty of our approach lies not in the extraction itself, but in leveraging these features as control targets within a generative framework for volatility surfaces. 

\subsection{Quantification Procedure} \label{sec:comp_feature}

Because the anchor point does not exist as a directly observable location on an IVS, we approximate the limiting features through a regression algorithm inspired by \citep{ait2021implied}.

First, each observed IVS sample $\mathbf{x}_i$ is smoothed by fitting a regression function over the strike--maturity grid:
\begin{equation}\label{eq:g}
    \hat{\sigma}^{\text{data}}_i(\tau, m) = g(\tau, m \mid \mathbf{x}_i) + \epsilon_i,
\end{equation}
where $m \in \mathbb{R}$, $\tau > 0$, and $\epsilon_i \sim \mathcal{N}(0, \sigma_y^2)$ represents observation noise. We can perform a bivariate Taylor expansion of $g$ around at the anchor point and estimate the coefficients using polynomial regression (see \autoref{sec:anchor} for details).  

Second, from the fitted regression, we extract feature values $\mathbf{y}_i$ by applying a differential operator $\mathbf{P}$,
\begin{equation} \label{eq:y-regression}
    \mathbf{y}_i = \mathbf{P} g(\tau, m \mid \mathbf{x}_i) + \epsilon_i,
\end{equation}
where $\mathbf{P}$ specifies which partial derivatives are evaluated at the anchor point. For non-negative integers $i,j$, define 
\begin{equation}
    \Sigma_{i,j} = \lim_{\tau \to 0^+} \frac{\partial^{i+j} g}{\partial \tau^i \, \partial m^j}(\tau,0),
\end{equation}
The controlled feature set is then
\begin{equation}
    \mathbf{G} \subseteq \{ \Sigma_{0,0}, \, \Sigma_{0,1}, \, \Sigma_{0,2}, \, \Sigma_{1,0} \},
\end{equation}
corresponding respectively to the volatility level, slope, curvature, and term-structure slope.

\paragraph{Remark.} In real market data, the ATM slope may explode to infinity for near-expiry options (time to maturity $\tau$ approaching zero). From the perspective of stochastic modeling, this short-end blow-up of the slope requires, for example,  incorporating  memory effects in volatility process  through fractional Brownian motions \citep{Bayer02062016}. In our current  approach, the polynomial regression can not deal with infinite values to capture such singularities; A possible solution is to adapt the basis functions to handle infinite slopes, which is left for future research.

\section{Controllable generation with VAE}\label{sec:CVAE}

In the context of machine learning, generative models aim to create realistic data samples by approximating an unknown and often intractable underlying distribution. Let $\mathbf{x} \in \mathcal{X} \subset \mathbb{R}^D$ denote an observed data sample (e.g., an implied volatility surface), where $\mathbb{R}$ denotes the real number space and $D$ is an integer. In practice, the underlying true distribution is inaccessible, and we only observe a finite set of independent and identically distributed samples $\{\mathbf{x}_i\}_{i=1}^n$.  Generative modeling seeks to learn a mapping from a simple latent distribution $\mathcal{Z} \subset \mathbb{R}^d$, for instance, a multivariate Gaussian distribution, to the complex high-dimensional data space $\mathcal{X}$.  New synthetic samples $\mathbf{x} \in \mathcal{X}$ are generated by sampling latent varaibles $\mathbf{z} \sim \mathcal{Z}$ and applying the learned mapping function to achieve $\mathbf{x} \approx g_\theta(\mathbf{z})$. In the VAE setting,  the latent space is much smaller than the data space, i.e., $0 < d \ll D$, and $g_\theta(\mathbf{z})$ represents a function approximated by a neural network with hidden parameters $\theta$ (e.g., hidden weights and biases).

To enable a controllable generation process, we extend the classical VAE framework \citep{kingma2014semi} by adding a set of controllable variables, in the spirit of conditional VAE. A related line of work on conditional generations with VAE involves semi-supervised VAEs \citep{kingma2014semi}, which embed label constraints to improve interpretability and control. Extensions such as \citep{li2019disentangled} and \citep{joycapturing} focus on disentangled representation learning, while \citep{paige2017learning} generalize the semi-supervised framework to continuous labels and dynamic dependencies among latents. Our approach draws on these ideas but tailors them to financial applications by combining interpretable latent design with quantifiable feature extraction for IVSs.

\subsection{Controllable generation of IVSs}

In our controllable VAE architecture, the generative process involves three types of variables.  $\mathbf{x}$ denotes the  data samples of IVSs, while  $\mathbf{y}$ represents a vector of controllable shape features, which can be extracted using the method in \autoref{sec:ShapeFeatures}.  The variable $\mathbf{z}$ denotes residual latent factors that capture IVS additional variability not covered by $\mathbf{y}$.  Together, the pair $(\mathbf{y},\mathbf{z})$ provides a disentangled latent representation in which $\mathbf{y}$ governs specified shape characteristics, and $\mathbf{z}$ preserves remaining structural variation of IVSs.

\begin{figure}[htbp]
    \centering
    \includegraphics[width=1\linewidth]{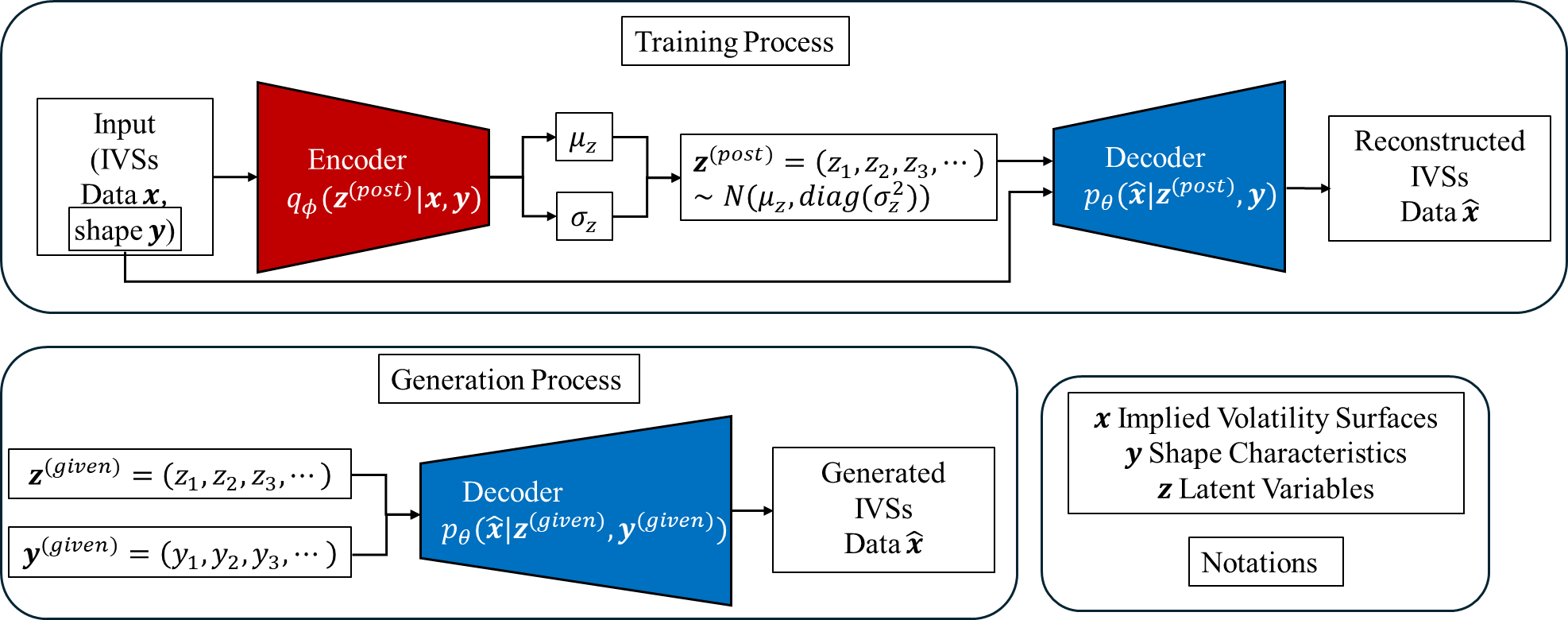}
    \caption{Flowchart of the controllable VAE framework for IVS generation.}
    \label{fig:CVAE-framework}
\end{figure}

\autoref{fig:CVAE-framework} outlines the overall workflow of the proposed controllable VAE. The training stage consists of an encoder and decoder network.  The encoder takes $(\mathbf{x},\mathbf{y})$ (an observed IVS and corresponding shape features) and maps them to the latent variable $\mathbf{z}$, while the decoder reconstructs $\mathbf{x}$ from the control/latent variables $(\mathbf{y},\mathbf{z})$.  Once training is complete, the generation stage relies solely on the decoder. By providing specific values of $y$ together with random samples of $z$ from the prior distribution, new IVSs can be produced that adhere to the desired feature specifications while maintaining diversity.

\begin{figure}[htbp]
    \centering
    \includegraphics[width=1\linewidth]{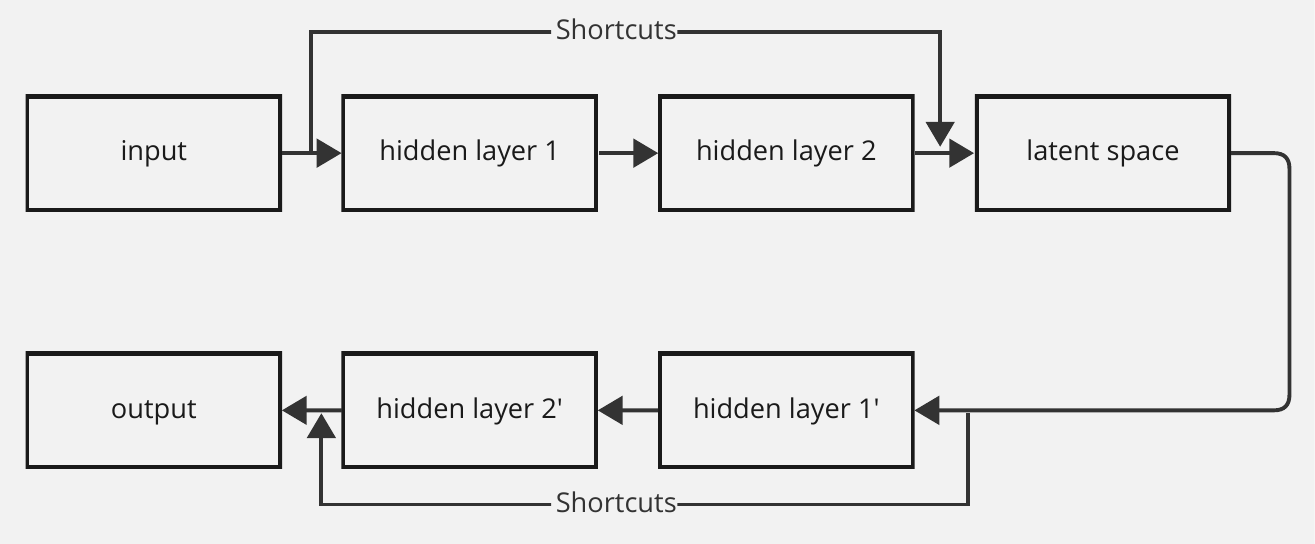}
    \caption{Neural network architecture with shortcut connections used in the controllable VAE.}
    \label{fig:NN_shortcuts}
\end{figure}

We employs a residual neural network (ResNet) for both the encoder and decoder, as shown in \autoref{fig:NN_shortcuts}. These architecture incorporates additional shortcut layers, which apply a linear transformation to the input or latent variables, allowing them to skip several layers and align with their dimensions. The transformed inputs are then added element-wise to the original outputs. We empirically find that the ResNet significantly improves the generative model performance, compared with plain feedforward architectures, which aligns with the findings in~\cite{he2015deepresiduallearningimage}.

\subsection{Generative process}

The controllable VAE framework extends the standard generative modeling setting 
by explicitly incorporating user-specified shape features. Let 
$\mathbf{x} \in \mathcal{X} \subset \mathbb{R}^D$ denote an IVS, $\mathbf{y}$ the 
vector of controllable shape characteristics extracted as described in 
Section~\ref{sec:ShapeFeatures}, and $\mathbf{z}$ the residual latent variables capturing 
additional variability not explained by $\mathbf{y}$. Generation then amounts to 
sampling $(\mathbf{y}, \mathbf{z})$ and decoding into an IVS.

We assume the following factorization,
\begin{equation}\label{eq:pxyz}
    p(\mathbf{x}, \mathbf{y}, \mathbf{z}) = p(\mathbf{z}) \, p(\mathbf{y}) \, 
    p_\theta(\mathbf{x} \mid \mathbf{y}, \mathbf{z}),
\end{equation}
where $p(\mathbf{z})$ and $p(\mathbf{y})$ are independent priors, and 
$p_\theta(\mathbf{x} \mid \mathbf{y}, \mathbf{z})$ is parameterized by the decoder 
network with parameters $\theta$. This independence assumption encourages 
a disentangled latent representation,i.e., $\mathbf{y}$ governs specific, interpretable 
shape features, while $\mathbf{z}$ accounts for residual variation.Notably, the dimensionality of $\mathbf{y}$ and $\mathbf{z}$ is typically much smaller than that of $\mathbf{x}$.  In this way, 
user-specified financial attributes (e.g., slope, curvature) can be imposed directly 
through $\mathbf{y}$, while diversity of surfaces is preserved through $\mathbf{z}$.

\subsection{Variational inference and training}
\label{sec:vae-training}

Training VAE aims to find optimal parameters $(\theta, \phi)$ such that the model distribution 
$p_\theta(\mathbf{x})$ approximates the true data-generating distribution. As in 
standard VAE formulations, this is achieved by maximizing the evidence lower 
bound (ELBO). For each  pair $(\mathbf{x}, \mathbf{y})$, we introduce an 
approximate posterior $q_\phi(\mathbf{z} \mid \mathbf{x}, \mathbf{y})$ and maximize
\begin{equation}
    \log p_\theta(\mathbf{x}, \mathbf{y}) \; \geq \; 
    \mathbb{E}_{\mathbf{z} \sim q_\phi(\mathbf{z}\mid \mathbf{x}, \mathbf{y})}
    \Big[ \log p_\theta(\mathbf{x} \mid \mathbf{y}, \mathbf{z}) \Big]
    - \beta \, \mathrm{KL}\!\left( q_\phi(\mathbf{z}\mid \mathbf{x}, \mathbf{y}) 
    \, \Vert \, p(\mathbf{z}) \right),
    \label{eq:vae-elbo}
\end{equation}
where $\beta$ is a trade-off parameter that balances reconstruction fidelity and 
latent regularization. The first term corresponds to reconstruction error, 
which under Gaussian assumptions reduces to mean squared error (see \autoref{sec:GaussianLoss}), while the second term regularizes the posterior to remain close to the prior distribution $p(\mathbf{z})$.

The training objective is therefore

\begin{equation}
\arg\min_{\theta, \phi} \mathcal{L}(\mathbf{x}, \mathbf{y}),
\end{equation}
where 
\begin{equation}\label{eq:Lxy}
    \mathcal{L}(\mathbf{x}, \mathbf{y}) \;=\; 
    - \mathbb{E}_{\mathbf{z} \sim q_\phi(\mathbf{z}\mid \mathbf{x}, \mathbf{y})} 
    \left[ \log p_\theta(\mathbf{x} \mid \mathbf{y}, \mathbf{z}) \right] 
    + \beta \, \mathrm{KL}\!\left(q_\phi(\mathbf{z}\mid \mathbf{x}, \mathbf{y}) 
    \Vert p(\mathbf{z}) \right).
\end{equation}
Training is performed to minimize the overall loss with respect to both encoder and decoder parameters $(\theta, \phi)$  by stochastic gradient descent optimization algorithms. 
Algorithmic details, including the full training loop, are provided in  Algorithm~\ref{algo:DVAE}

\subsubsection{The reparametrization trick}
While the principles of VAEs do not impose any specific assumptions on the underlying variables, practical implementation typically requires certain assumptions to ensure tractability. For tractability, the residual latent variable is assumed to follow the standard Gaussian distribution,  
\begin{equation}\label{eq:prior}
\mathbf{z} \sim \mathcal{N}(0, \mathbf{I}),
\end{equation}
and the posteriors have the following Gaussian form  
\begin{subequations}\label{eq:posterior}
    \begin{align}
        q_{\phi}(\mathbf{z} \mid \mathbf{x}, \mathbf{y}) &\sim \mathcal{N} \big(\mathbf{z} \mid \boldsymbol{\mu}_{z},\, \mathrm{diag}(\boldsymbol{\sigma}^2_{z}) \big),\\
        p_{\theta}(\mathbf{x} \mid \mathbf{y, z}) &\sim \mathcal{N}\big(\boldsymbol{\mu}_{x},\, \mathbf{I}\big).
    \end{align}
\end{subequations}
These assumptions yield closed-form expressions for the KL term in the objective function. 

Regarding the controllable variable $\mathbf{y}$, we assume the regression error in Equation~\eqref{eq:y-regression} is negligible when extracting shape features  (Section~\ref{sec:ShapeFeatures}). 

In such case, the variable $\mathbf{y} \mid \mathbf{x}$ becomes deterministic, and we have the posterior
\begin{equation} \label{eq:yonx}
q(\mathbf{y} \mid \mathbf{x}) = \delta(\mathbf{y}^{\text{obs}}(\mathbf{x})),
\end{equation}
with  $\delta(\cdot)$ being Dirac delta function.
During training,  the input of the decoder consists of two conditional variables, $\mathbf{z} \mid (\mathbf{x},\mathbf{y}) $ parametrized by the encoder and $\mathbf{y} \mid \mathbf{x} $ determined by the function of computing  IVS shape features.  In  our setting, $\mathbf{z} \mid (\mathbf{x},\mathbf{y}) $ is a random variable,  while $\mathbf{y} \mid \mathbf{x} $ is a deterministic one. In other words,  the residual variable $\mathbf{z}$ requires sampling, while the controllable variable $\mathbf{y}$ does not.

The sampling process is inherently non-differentiable, which prevents gradients from being directly propagated through the neural networks. To enable back-propagation in the stage of training,  a popular effective reparametrization trick reads,
\begin{align}
(\boldsymbol{\mu}_z, \log \boldsymbol{\sigma}_z) &= \mathrm{EncoderNN}_{\phi}(\mathbf{x}, \mathbf{y}), \\
\mathbf{z} &= \boldsymbol{\mu}_z + \boldsymbol{\sigma}_z \odot \boldsymbol{\epsilon},
\end{align}
which allows stochastic sampling to be expressed as a differentiable transformation. This technique enables efficient training of the encoder–decoder network while preserving stochasticity in the latent representation.

\subsection{Post-generation correction for no-arbitrage}\label{sec:repair_method}

Although the controllable VAE introduced above allows flexible and interpretable 
generation of IVSs, the outputs are not automatically guaranteed to satisfy the 
 no-arbitrage constraints described in Section~\ref{sec:arbitrage_free_conditions}. 
Violations may occur in particular when latent variables $\mathbf{z}$ are sampled from 
regions far from the in-sample manifold  determined by the training dataset, for example,  in the tails of the Gaussian 
distribution. This is because the training objective does not directly 
enforce arbitrage-free conditions.

There are two possible ways to mitigate this issue: (i) resampling $\mathbf{z}$ until a valid surface is generated, or (ii) refining $\mathbf{z}$ through optimization. We adopt the latter and propose a post-generation correction procedure that operates in the latent space. The basic idea is to adjust the residual variables $\mathbf{z}$ while keeping the controllable features $\mathbf{y}$ fixed, thereby ensuring the specified shape characteristics are preserved as much as possible. By optimizing in the latent space rather than directly on the surface, the adjustment stays within the class of volatility surfaces that the decoder can generate, which avoids ad hoc modifications that might break controllability or introduce unrealistic shapes. Moreover, to prevent large deviations, a proximity term is included so that the adjusted IVS remains close to the originally generated surface.

For an arbitrage-violating IVS decoded from $(\mathbf{y}, \mathbf{z})$,  we define a composite loss function that penalizes violations of both the calendar-spread and butterfly conditions, along with a regularization term that discourages excessive deviation from the original IVS:
\begin{equation}
    \mathcal{L}_{\text{total}} = \mathcal{L}_{\text{Calendar}} + \mathcal{L}_{\text{Butterfly}} + \mathcal{L}_{\text{MSE}}.
\end{equation}

Following \autoref{sec:arbitrage_free_conditions},  the two arbitrage penalty terms are defined as follows:

\begin{equation}
    \mathcal{L}_{\text{Calendar}} = \frac{1}{MM_{\tau}} \sum_{i=1}^{M} \sum_{j=1}^{M_\tau} \max\left(0, -\frac{\partial w}{\partial \tau}(m_i, \tau_j)\right),
\end{equation}

\begin{equation}
    \mathcal{L}_{\text{Butterfly}} = \frac{1}{M_\tau} \sum_{j=1}^{M_\tau} \left( \frac{1}{M} \sum_{i=1}^{M} \max\left(0, -f(m_i; \tau_j)\right) \right),
\end{equation}
where \( M \) and \( M_\tau \) denote the number of evaluation points along the \( m \)- and \( \tau \)-axes, respectively, and $f(m; \tau_j)$ is defined as Equation~\eqref{eq:butterfly}. The third term, \( \mathcal{L}_{\text{MSE}} \), measures the mean squared error between the original (arbitrage-violating) IVS and the adjusted (arbitrage-repaired) IVS.

The post-generation correction method 
refines $\mathbf{z}$ by minimizing the loss function $\mathcal{L}_{\text{total}}$, while the value of  $\mathbf{y}$ stays the same as that of the problematic IVS. We employ the L-BFGS optimizer~\cite{liu1989limited} to minimize 
$\mathcal{L}_{\text{total}}$, terminating when either the gradient norm or the total loss 
 falls below a specified threshold. The implementation details are 
summarized in Algorithm~\ref{algo:ArbFree}.

\paragraph{Remark.}
The L-BFGS method is a local optimization scheme, which makes it prone to convergence toward local minima rather than the global optimum. The quality and stability of the obtained solutions can vary, depending on practical design factors such as the choice of the initial guess, the termination threshold, and other implementation parameters. As a result, the optimization outcome may differ across runs.

\section{Numerical Results}\label{sec:Numerical}

This section is to evaluate  the proposed controllable generative modelling framework through  numerical experiments: (i) single-feature control, (ii) simultaneous control of multiple features, and (iii) arbitrage-free validation and repair. To provide a systematic comparison, we report both quantitative error metrics and representative visualizations.  The hyperparameters and training settings of the VAE are summarized in Table~\ref{tab:CVAE}.

An important design choice in a VAE is the dimensionality of the latent space. A high-dimensional latent space may introduce redundancy, with several latent variables encoding overlapping information, while a very low-dimensional space may fail to fully capture the relevant structure of the IVS data. To balance these considerations, we experimented with different dimensionalities and fixed the latent dimension at 5 in all experiments. Notably, increasing the number of latent dimensions beyond this value does not necessarily yield further improvement in model performance, as discussed in~\citep{ning2023arbitrage}.

The training dataset consists of 60,000 IVSs, each represented on a fixed grid with log-moneyness $m \in [-0.27, 0.27]$ and time to maturity $\tau \in [0.1, 0.6]$, both discretized into 28 points. Additional details are provided in Appendix~\ref{sec:ParaModels}. To assess dataset independence, we trained the model on both a Heston dataset and a combined Heston–SABR dataset. The controllable VAE achieved similar controllability and accuracy in both cases, indicating robustness to dataset choice. The results only for the combined dataset are presented in the paper.

\begin{table}[htbp]
\centering
\caption{Hyperparameters used in the controllable VAE model.}
\label{tab:CVAE} 
\begin{tabular}{@{}lc@{}}
\toprule[1.2pt]
\textbf{Hyperparameter} & \textbf{Setting} \\
\midrule
Input Dimension ($\mathbf{x}$)             & 784 \\
Control Dimension ($\mathbf{y}$)           & problem-specific \\
Latent Dimension ($\mathbf{z}$)            & 5 \\
Encoder hidden layers                      & [256, 128] \\
Decoder hidden layers                      & [128, 256] \\
Batch size                                 & 64 \\
Activation function                        & ReLU \\
Optimizer                                  & Adam \\
Learning rate                              & $3 \times 10^{-4}$ \\
$\beta$                                    & 
problem-specific \\
Max epochs                                 & 5000 \\
\bottomrule[1.2pt]
\end{tabular}
\end{table}

\begin{figure}[htbp]
    \centering
    \includegraphics[width=1\linewidth]{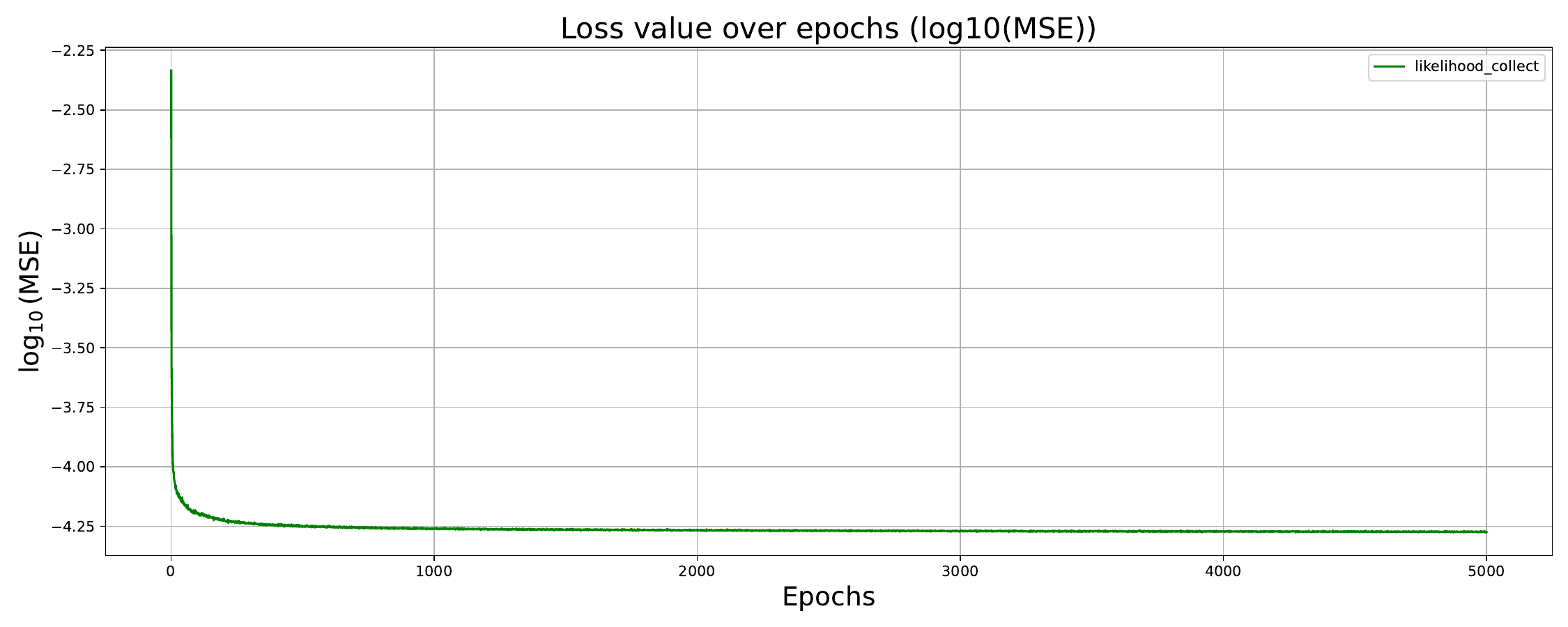}
    \caption{Convergence of the log-transformed MSE loss during training.}
    \label{fig:train_loss_log10MSE}
\end{figure}

During training, we monitor two loss metrics: the total loss (i.e., negative ELBO) and the reconstruction error which is measured by the mean squared error (MSE) between input and reconstructed IVSs. In Experiment~I, the total loss stabilizes by the end of training (5000 epochs)  to be -0.9825. \autoref{fig:train_loss_log10MSE} shows the corresponding reconstruction error, which converges to the order of around $-4$ (i.e., $\log_{10} (\text{MSE}) \approx -4$). Similar convergence patterns are observed across all of our experiments.

After training, model performance is assessed in three complementary ways. First, histograms of the differences between target and generated feature values quantify the overall controllability and precision. Second, controllability tests are performed by varying  $\mathbf{y}$ while fixing $\mathbf{z}$ to examine whether the decoder produces IVSs exhibiting the specified feature values. Third, latent traversals are conducted by varying  $\mathbf{z}$ while keeping $\mathbf{y}$ fixed to examine whether non-targeted variations can be captured without altering the controlled features. Together, these evaluations provide a comprehensive assessment of accuracy, controllability, and generative flexibility.

\subsection{Experiment I: Controlling a Single Feature}
\label{sec:exp1}

We begin with controlling a single feature, i.e., the volatility level $y_L$, as it is the most fundamental shape characteristic of IVSs and closely linked to market stress scenarios (e.g., recessions with elevated volatility levels). By setting $\mathbf{y}=(y_L), \mathbf{z}=(z_1,z_2,z_3,z_4, z_5)$, the objective is to verify whether the controllable VAE can precisely manipulate this individual feature, while other features are captured by the latent variables.

\subsubsection{Overall Performance}\label{sec:OverallPerformance}
We generated 1,000 IVSs by specifying random values for $y_L$,  meanwhile sampling $\mathbf{z}$ from the standard normal distribution $\mathcal{N}(\mathbf{0}, \mathbf{I})$. 
For each case, we define the generation error
$$
e_L = y_{L}^{\text{(generated)}} - y_{L}^{\text{(given)}},
$$
whered $y_{L}^{\text{(given)}}$ stands for the desired level which is given to the input of the generator (i.e., the already trained decoder),  and $y_{L}^{\text{(generated)}}$ stands for the volatility level of a generated IVS, which is  computed using the regression method in Section~\ref{sec:comp_feature}.

\begin{figure}[htbp]
    \centering
    \includegraphics[width=.7\linewidth]{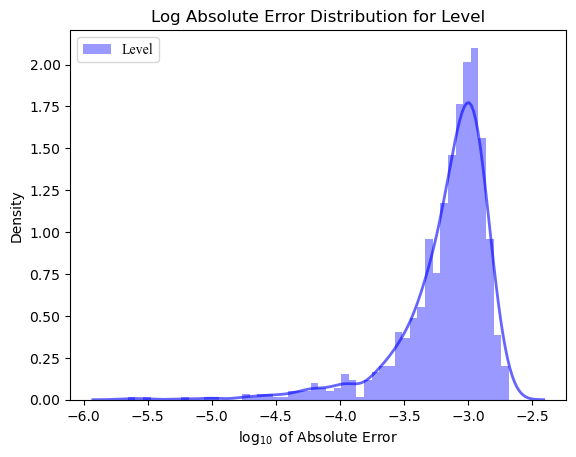}
    \caption{Distribution of the log absolute errors $\log_{10} |e_L|$ between user-specified  and  model-generated IVS levels.}
    \label{fig:histLevel}
\end{figure}

We measure the control accuracy using the absolute error $|e_L|$. The distribution of $\log_{10} |e_L|$ across the 1,000 samples is shown in \autoref{fig:histLevel}. In most cases, the generation error $|e_L|$ falls below $0.1\%$, while the maximum deviation is reflected by $\max(\log_{10} |e_L|) \approx -2.1$. Furthermore, there are no calendar/butterfly violations within these generated IVSs. These results demonstrate that the controllable VAE achieves high numerical precision in reproducing the target \textit{level} feature.

\subsubsection{Varying $y_L$ while freezing $\mathbf{z}$}\label{sec:single_control_varyY}
As a complement to the aggregate results in Section~\ref{sec:OverallPerformance}, this section takes a closer look at two representative cases. Our aim here is to examine how the generator realizes prescribed  volatility levels $y_L$ at fixed $\mathbf z$.

Given a specific level $y_L$, multiple volatility surfaces can exist with different shapes; this variation is represented by the latent vector $\mathbf{z}$. Two latent settings are shown to illustrate the effect of changing $y_L$ at fixed $\mathbf z$: $\mathbf{z}=(0,0,0,0,0)$ (mean region in latent) and $\mathbf{z}=(0,-6,0,0,0)$ (Gaussian tail in latent). The latter one, a sample of the $6\sigma$ domain along the second latent variable $z_2$ under the standard normal prior, is included to probe robustness in a low-density part of the latent space. We plot 2D slices of implied volatility versus log-moneyness $m$ at $\tau\in\{0.10,0.35,0.60\}$, which are the first, middle, and last points of the 28-value maturity grid, respectively. In each plot, the red cross marks $(m=0,\tau=0.10)$, the existing grid point closest to the anchor point $(m=0,\tau\to0^+)$. 

\begin{figure}[htbp]
    \centering
    \includegraphics[width=1\linewidth]{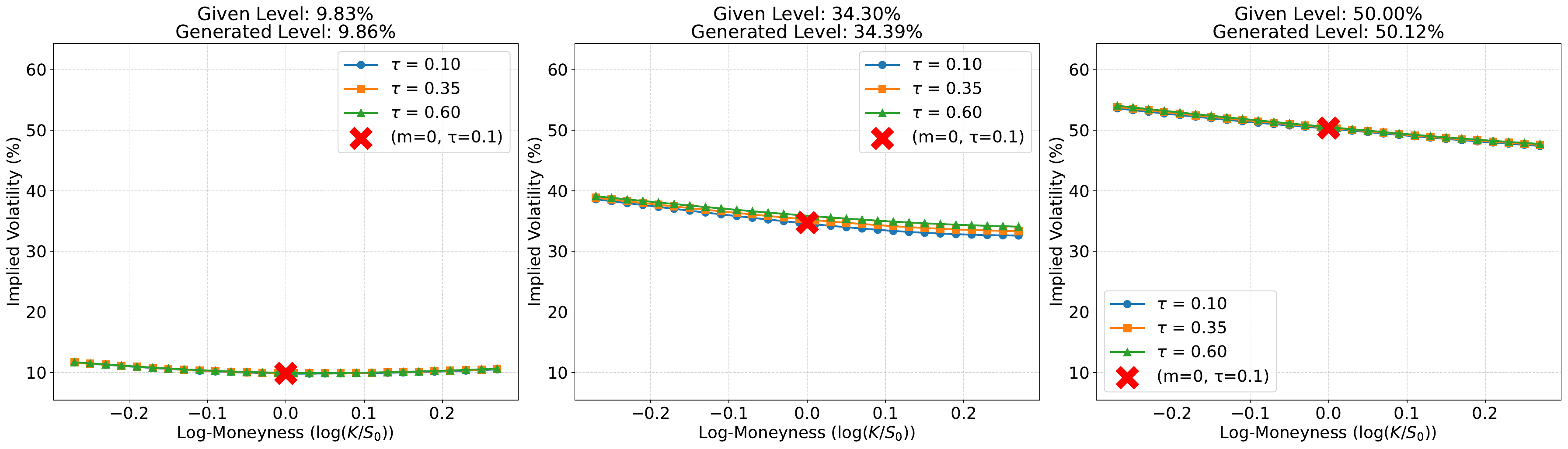}
    \caption{From left to right, given/generated $y_L=9.83/9.86,34.30/34.39,50.00/50.12(\%)$, while $\mathbf{z}=(0,0,0,0,0)$.}
    \label{fig:2d_Level_HS_min_mean_max}
\end{figure}

\begin{figure}[htbp]
    \centering
    \includegraphics[width=\linewidth]{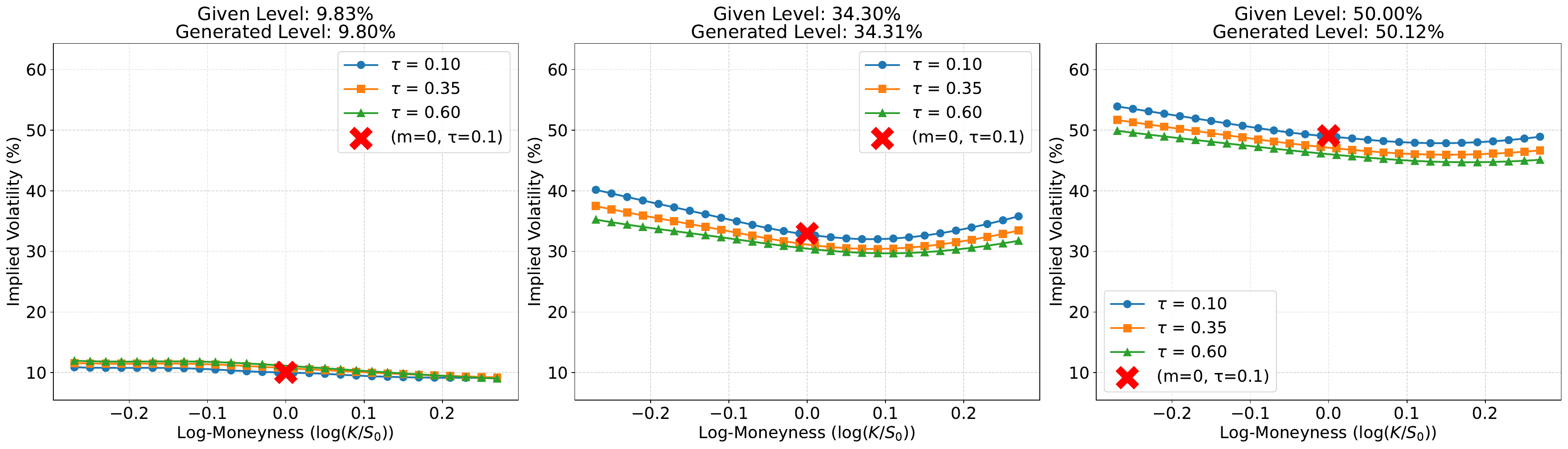}
    \caption{From left to right, given/generated $y_L=9.83/9.83,34.30/34.31,50.00/50.12(\%)$, while $\mathbf{z}=(0,-6,0,0,0)$}.
    \label{fig:2d_Level_smile_min_mean_max}
\end{figure}

The first case, shown in \autoref{fig:2d_Level_HS_min_mean_max}, produces a series of relatively flat IVSs under the specified volatility level values. When varying $y_L$, the generator produces different IVSs to align with the desired feature. As $y_L$ changes from $9.83\%$ to $34.30\%$ to $50.00\%$, the generated levels match the targets with absolute errors of $0.03\%$, $0.09\%$, and $0.12\%$, respectively, and relative errors (i.e., $|e_L|/y_L^{(\text{given})}$) of $0.31\%$, $0.26\%$, and $0.24\%$, respectively. The second case, shown in \autoref{fig:2d_Level_smile_min_mean_max}, produces a series of smile-shaped IVSs under the same volatility level values. Similar to the first case, the generated levels match the targets with small absolute errors of $0.03\%$, $0.01\%$, and $0.12\%$, respectively, and relative errors of $0.31\%$, $0.03\%$, and $0.24\%$, respectively.

Importantly, to satisfy arbitrage-free conditions, the generator does not simply shift the IVS. For example, in \autoref{fig:2d_Level_smile_min_mean_max}, the left IVS is flat, while the middle one exhibits a pronounced smile pattern as  $y^{(\text{given})}_L$ varies from $9.83\%$ to $34.30\%$. A naive shift of an implied-volatility surface (e.g., adding a constant to the implied volatility everywhere) does not, in general, preserve static no-arbitrage constraints. This can be explained by the no-calendar-arbitrage condition, which requires that for each fixed moneyness $m$, the total implied variance $w(m,\tau)=\sigma^2(m,\tau),\tau$ is non-decreasing in maturity $\tau$, i.e., $\partial_{\tau} w(m,\tau)\ge 0$. A vertical volatility shift $\sigma_{\text{new}}(m,\tau)=\sigma(m,\tau)+\delta$ gives $w_{\text{new}}(m,\tau)=(\sigma(m,\tau)+\delta)^2\tau$, and 
$$\partial_\tau w_{\text{new}}(m,\tau)=(\sigma(m,\tau)+\delta)^2 + 2(\sigma(m,\tau)+\delta)\,\tau\,\partial_\tau\sigma(m,\tau),$$
which may become negative where $\partial_\tau\sigma(m,\tau)$ is sufficiently negative, violating no-calendar-arbitrage.

Overall, the generator achieves stable and accurate control over the \textit{level} shape feature while producing realistic diverse IVSs.

\subsubsection{Varying $\mathbf{z}$ while freezing $y_L$}\label{sec:varyZ}
In this section, we fix the controllable feature $y_L$ and  vary the latent variables $\mathbf{z}$ to investigate whether the latent space implicitly encodes other hidden characteristics of the IVS. 

\begin{figure}[htbp]
    \centering
    \includegraphics[width=1\linewidth]{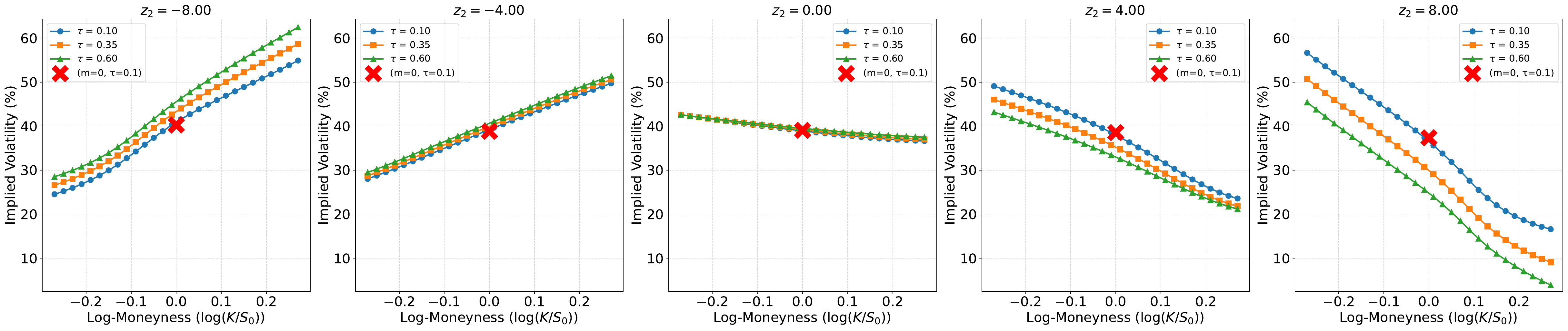}
    \caption{Given $\mathbf{y} = (y_L) = (0.3875)$, the figure shows the effect of varying $z_2 = -8, -4, 0, 4, 8$ from left to right, while keeping all other components of $\mathbf{z}$ fixed at zero. The corresponding slope values from left to right are $0.5792$, $0.4078$, $-0.1166$, $-0.5211$, and $-0.8095$.}
    \label{fig:z2_ivs_2d_surface}
\end{figure}

\autoref{fig:z2_ivs_2d_surface} shows five IVSs generated by fixing the level at $38.85\%$ and varying $z_2$ uniformly from $-8$ to $8$, while setting the other latent variables to be zero. It is visually evident that $z_2$ primarily modulates the \textit{slope} of the IVS, although some mild entanglement with the \textit{term structure} and \textit{curvature} features is also observed. The two-dimensional slices along the log-moneyness direction, reveals a clear monotonic decrease in slope across the five IVS instances. The corresponding slope values from left to right are $0.5792$, $0.4078$, $-0.1166$, $-0.5211$, and $-0.8095$, indicating a strong and consistent correlation between the slope feature and the latent dimension $z_2$, with Pearson coefficient\footnote{The Pearson correlation coefficient between two variables $\mathbf{a} = [a_1, \dots, a_n]$ and $\mathbf{b} = [b_1, \dots, b_n]$ is defined as
\[
r = \frac{\sum_{i=1}^{n} (a_i - \bar{a})(b_i - \bar{b})}{\sqrt{\sum_{i=1}^{n} (a_i - \bar{a})^2} \sqrt{\sum_{i=1}^{n} (b_i - \bar{b})^2}},
\]
where $\bar{a}$ and $\bar{b}$ denote the sample means of $\mathbf{a}$ and $\mathbf{b}$, respectively. The value of $r$ ranges from $-1$ (perfect negative linear correlation) to $1$ (perfect positive linear correlation), with $0$ indicating no linear correlation.} approximately $-0.991$. Notably, in our dataset, the observed range of slope values is between $-0.5642$ and $0.3381$. In this example, both the upper and lower slope values exceed those bounds, suggesting that manipulating $z_2$ can produce IVSs with slope values beyond the training data range. This implies that the model not only learns the underlying representation of slope but can also extrapolate meaningfully in the latent space. 

\begin{figure}[htbp]
    \centering
    \includegraphics[width=1\linewidth]{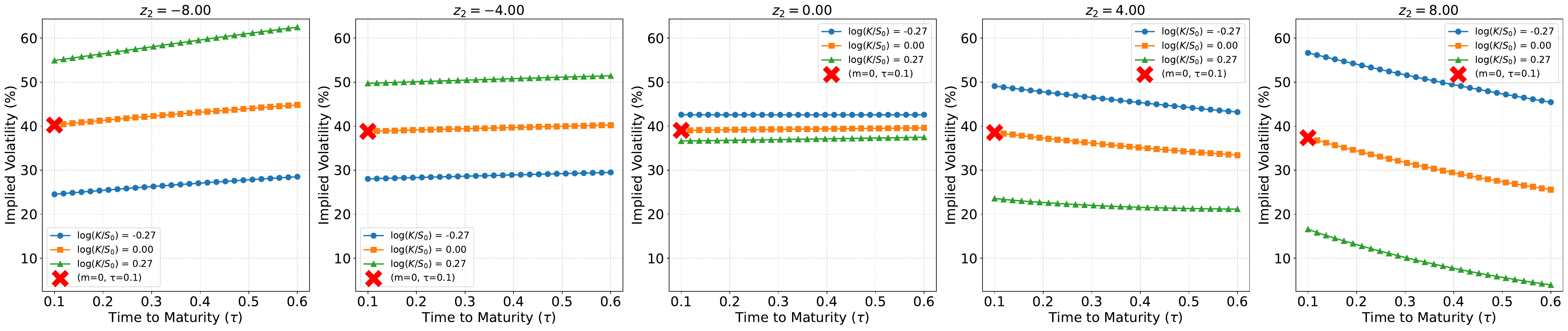}
     \caption{Given $\mathbf{y} = (y_L) = (0.3875)$, the figure shows the effect of varying $z_2 = -8, -4, 0, 4, 8$ from left to right, while keeping all other components of $\mathbf{z}$ fixed at zero. The corresponding term structure values from left to right are $0.1245$, $0.0341$, $0.0097$, $-0.1351$, and $-0.3368$.}
    \label{fig:z2_ivs_2d_term}
\end{figure}

\autoref{fig:z2_ivs_2d_term} presents two-dimensional slices along the term-structure direction. The associated \textit{term structure} values from left to right are $0.1245$, $0.0341$, $0.0097$, $-0.1351$, and $-0.3368$, suggesting a secondary association with $z_2$, with Pearson coefficient being $-0.958$. 

The \textit{level} values across these five IVS instances remain tightly controlled around $38.85\%$, with deviations within $\pm 0.05\%$, again demonstrating the precise controllability of our method. In contrast, the \textit{curvature} values show a non-monotonic trend, peaking in the middle and reversing sign across the set, exhibiting no clear correspondence with $z_2$ with Pearson coefficient about $0.021$.

\begin{figure}[htbp]
    \centering
    \includegraphics[width=1\linewidth]{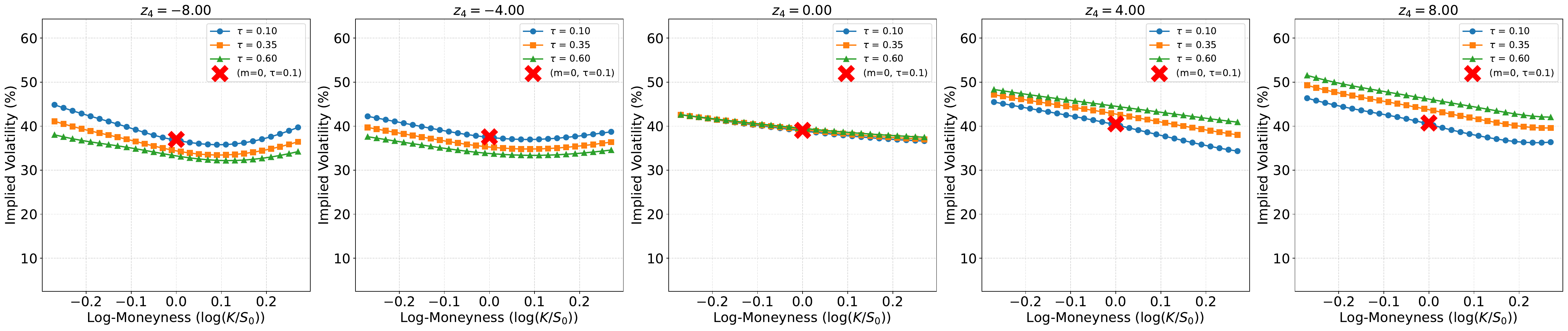}
    \caption{Given $\mathbf{y} = (y_L) = (0.3875)$, the figure shows the effect of varying $z_4 = -8, -4, 0, 4, 8$ from left to right, while keeping all other components of $\mathbf{z}$ fixed at zero. The corresponding curvature values from left to right are $1.1707$, $0.7628$, $0.1623$, $-0.0487$, and $0.1603$.}
    \label{fig:z4_ivs_2d_surface}
\end{figure}

\begin{figure}[htbp]
    \centering
    \includegraphics[width=1\linewidth]{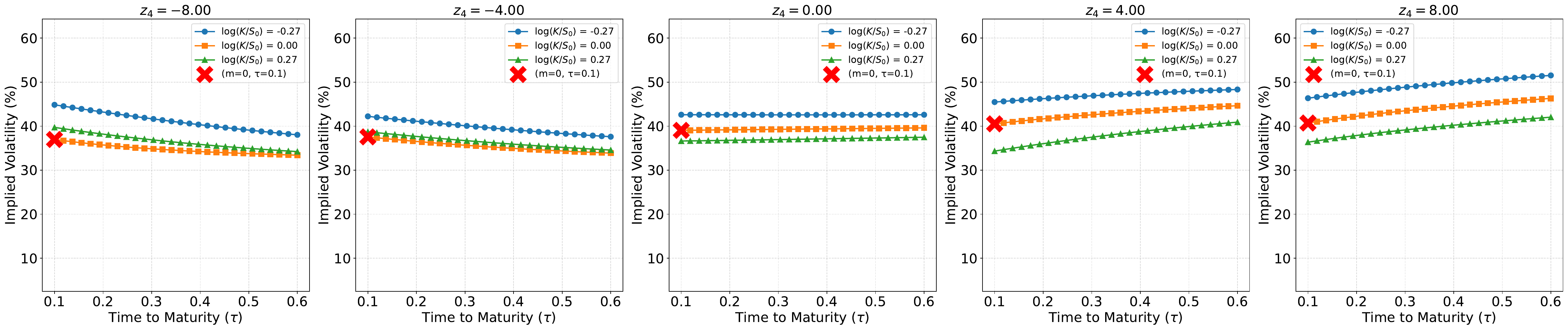}
    \caption{Given $\mathbf{y} = (y_L) = (0.3875)$, the figure shows the effect of varying $z_4 = -8, -4, 0, 4, 8$ from left to right, while keeping all other components of $\mathbf{z}$ fixed at zero. The corresponding term structure values from left to right are $-0.1557$, $-0.1221$, $0.0097$, $0.1355$, and $0.1717$.}
    \label{fig:z4_ivs_2d_term}
\end{figure}

Similarly, latent dimension $z_4$ primarily captures a combination of \textit{curvature} and \textit{term structure} characteristics, as illustrated in the three-dimensional plots in \autoref{fig:z4_ivs_effect}, the log-moneyness slice in \autoref{fig:z4_ivs_2d_surface}, and the term-structure slice in \autoref{fig:z4_ivs_2d_term}.

However, associations inferred from the latent space do not always admit a consistent quantitative interpretation: directions that should increase the slope sometimes appear to decrease it in visualizations, and vice versa.

In summary, the controllable VAE generates IVSs with high fidelity for the targeted level feature, while residual latent variables preserve shape diversity.   The latent space may appear to disentangle slope and curvature but not in a quantitatively reliable way, reinforcing the need for explicit controllable variables.

\subsection{Experiment II: Controlling Multiple Shape Features}
\label{sec:exp2}

 We next investigate the ability of the controllable VAE  to simultaneously generate multiple desired IVS shape features. The objectives are twofold: (i) to test stability and accuracy as the dimensionality of controllable features increases, and (ii) to evaluate whether the residual latent variable $\mathbf{z}$ continues to capture IVS hidden structures when more shape characteristics are explicitly supervised.

We begin by controlling three features: the \textit{level} ($y_L$), \textit{slope} ($y_S$), and \textit{term structure} ($y_T$).  A  extreme scenario is followed by  including the \textit{curvature} component ($y_C$) to test four-dimensional control, yielding the  vector $\mathbf{y} = (y_L, y_S, y_C, y_T)$, which accounts for most of the  variance in our IVS dataset. 

\paragraph{Remark.} \label{para:beta}
The hyperparameter $\beta$ controls the relative weighting of the latent variable $\mathbf{z}$ term in the objective function. In one-, three-, and four-feature control experiments, we observe that smaller values of $\beta$ enable the model to more effectively capture residual shape features. when $\beta$ is too large, the features encoded in the latent space tend to become increasingly entangled with the explicitly controlled features $\mathbf{y}$.

\subsubsection{Three shape features} \label{sec:threefeatures}
 In the three-feature setting,  we control $y=(y_L, y_S, y_T)$ and keep the training hyperparameters of Experiment~I, using $\beta=5\times10^{-8}$.

\begin{figure}[htbp]
    \centering
    \includegraphics[width=.8\linewidth]{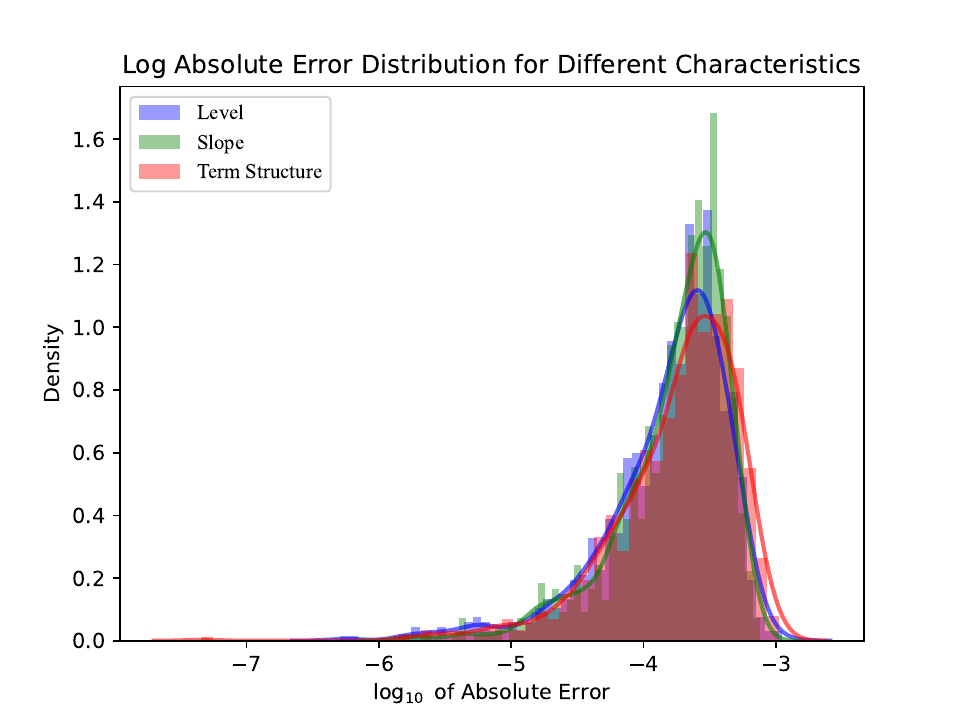}
    \caption{Distribution of log absolute errors for controlled level, slope, and term-structure features under the three-feature setting based on $1,000$ samples}.
    \label{fig:errorLST}
\end{figure}

\autoref{fig:errorLST} summarizes the  control accuracy over three features. 
Most absolute generation errors fall between  $10^{-5}$ and $10^{-3}$ (equivalently, $\log_{10} |e| \in [-5,-3]$). The maximum absolute errors are approximately $5.5\times 10^{-4}$ for level, $8\times 10^{-4}$  for the slope over moneyness, and $5.4\times 10^{-4}$ for the term structure slope. These results indicate the generator achieves precise control over the three IVS features together.  Compared to the single-feature control in \autoref{sec:exp1}, the controllable generative model for three features achieves comparable accuracy, as their generation errors are  of the similar order.

\begin{figure}[htbp]
    \centering
    \includegraphics[width=1\linewidth]{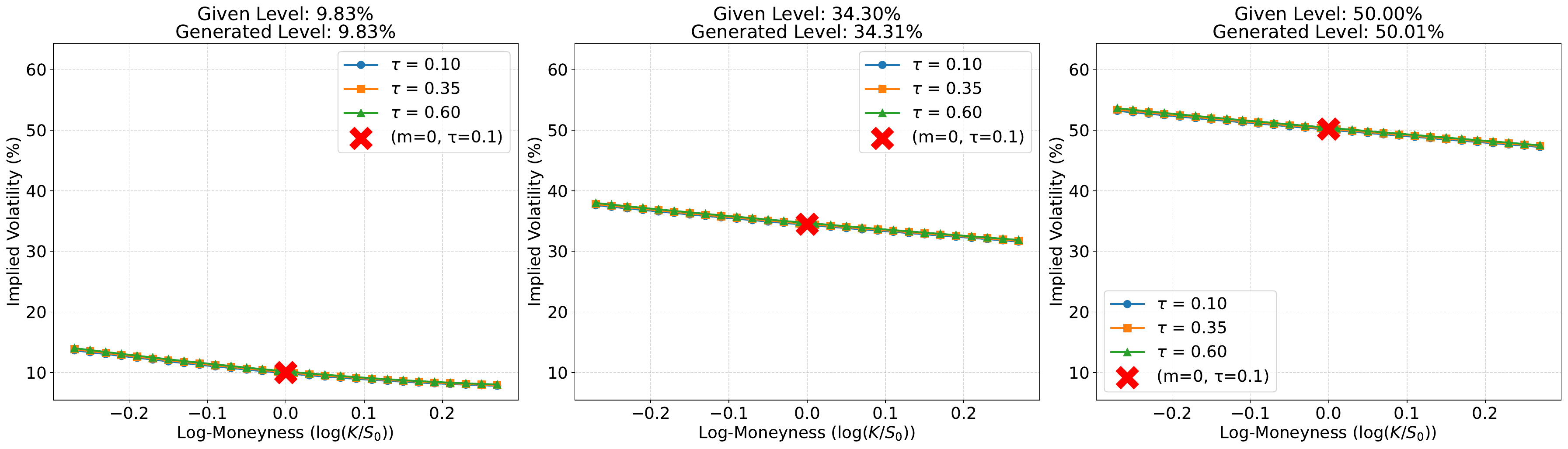}
    \caption{From left to right, given/generated $y_L=9.83/9.83,34.30/34.31,50.00/50.01(\%)$ with fixed $\mathbf{z}$.}
    \label{fig:2dl}
\end{figure}

\begin{figure}[htbp]
    \centering
    \includegraphics[width=1\linewidth]{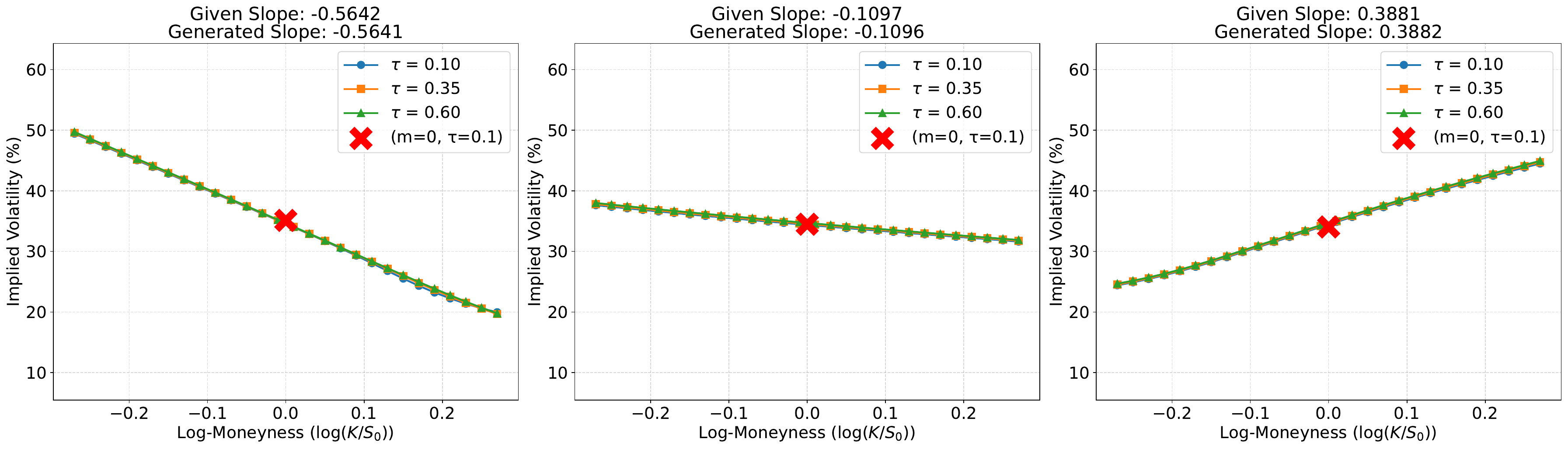}
    \caption{From left to right, given/generated $y_S=-0.5642/-0.5641,-0.1097/-0.1096,0.3881/0.3882$ with fixed $\mathbf{z}$.}
    \label{fig:2ds}
\end{figure}

\begin{figure}[htbp]
    \centering
    \includegraphics[width=1\linewidth]{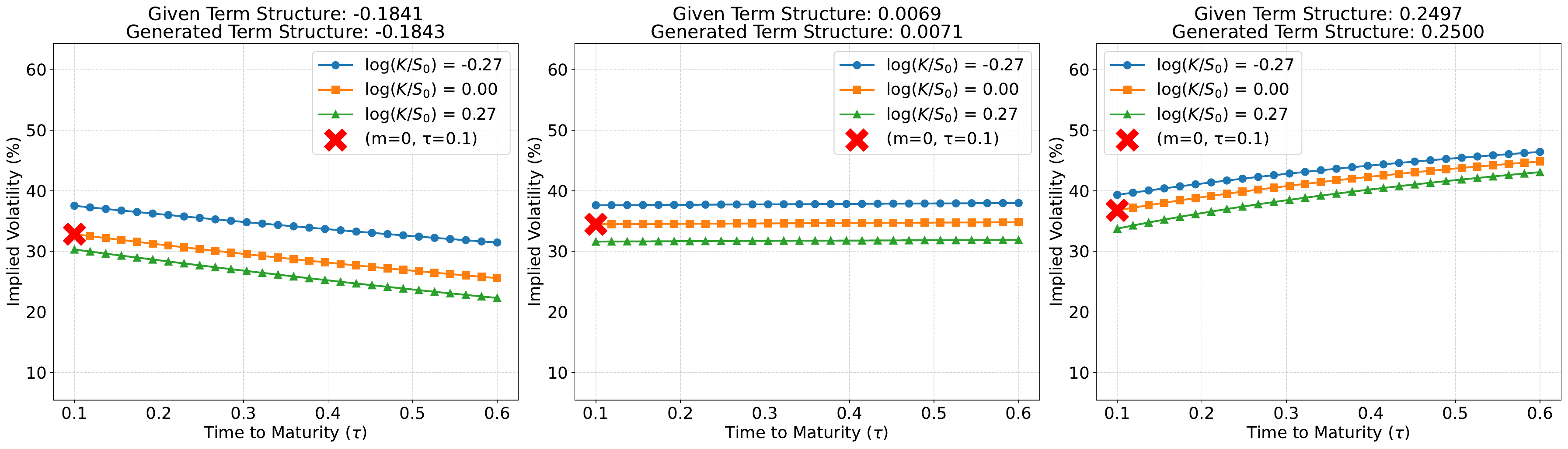}
    \caption{From left to right, given/generated $y_T=-0.1841/-0.1843,0.0069/0.0071,0.2497/0.2500$ with fixed $\mathbf{z}$.}
    \label{fig:2dt}
\end{figure}

To illustrate the effect of simultaneously specifying three features, we present 2D views (their 3D counterpart can be found in Appendix {\ref{sec:exp2}) highlighting how the generated IVSs respond when one of the targets is varied while the other two are kept fixed (together with $\mathbf{z}$).  
\begin{itemize}
  \item \textbf{Level} in \autoref{fig:2dl}. The volatility surfaces move vertically as intended, similar to Experiment I. Meanwhile, the slope along moneyness $y_S$ and term-structure slope $y_T$ stay unchanged.  
  \item \textbf{Slope } in \autoref{fig:2ds}. The tilt along moneyness adjusts to be negative or positive as specified,  with the level $y_L$ and term-structure slope $y_T$  preserved.  
  \item \textbf{Term-structure} in \autoref{fig:2dt}. The term-structure profile steepens or flattens accordingly, while the level $y_L$ and slope $y_S$ remain stable.  
\end{itemize}

These results indicate the generator achieves feature-wise disentanglement, i.e., when one component of $\mathbf{y}$ is varied, the generated IVSs reflect the desired adjustment in that specific feature, while the other controlled features remain  stable within negligible error margins.

\begin{figure}[htbp]
    \centering
    \includegraphics[width=1\linewidth]{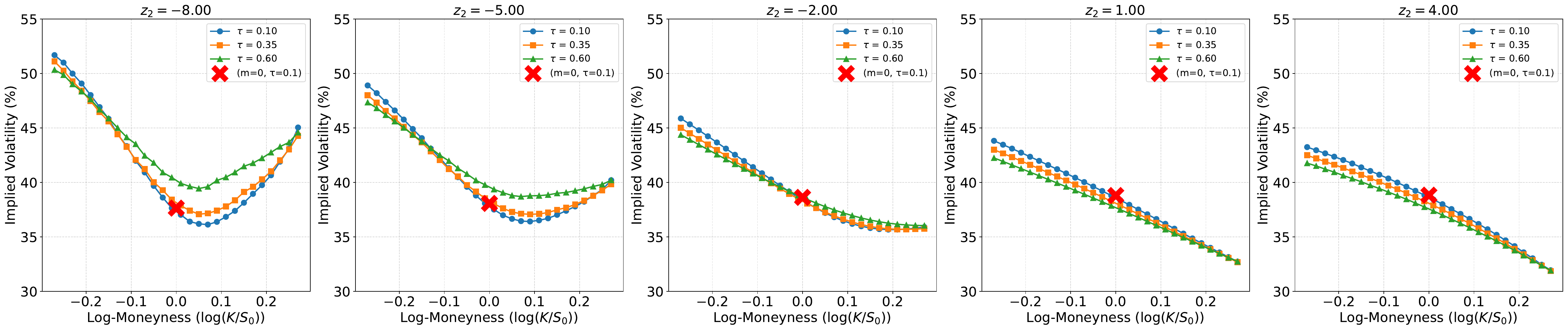}
     \caption{Given $\mathbf{y} = (y_L, y_S, y_T) = ( 0.3875, -0.2139, -0.0236)$, the figure shows the effect of varying $z_4 = -8, -5, -2, 1, 4$ from left to right, while keeping all other components of $\mathbf{z}$ fixed at zero. The corresponding curvature values from left to right are $2.5750$, $1.5353$, $0.5744$, $-0.0700$, and $-0.2504$.}
    \label{fig:2dz2LST}
\end{figure}

We then examine whether the latent variables still encode meaningful structures when three features are supervised. 
We find that the second latent component $z_2$ predominantly corresponds to a curvature-related effect. By varying $z_2$ from $-8$ to $4$, the curvature of the generated IVS decrease monotonically across five generated samples,  from left to right in~\autoref{fig:2dz2LST},  whose curvature values are, respectively, $2.5750$, $1.5353$, $0.5744$, $-0.0700$ and $-0.2504$. Importantly, despite the significant change in curvature, the  controlled features remain highly stable, that is, both the level and slope absolute error are smaller than $7\times10^{-4}$, and the term structure absolute error is less than $3\times10^{-4}$.

\subsubsection{Four shape features}
\label{sec:extremeResults}

In the previous experiments, the residual latent variables $\mathbf{z}$ still carried interpretable variation of IVSs. This raises an important question: what happens to the role of $\mathbf{z}$ once all four principal stylized factors involved in our training IVS dataset are explicitly supervised? To address this, we extend control to $y=(y_L,y_S,y_C,y_T)$, thereby covering the major stylized factors of IVSs. This experiment allows us to assess whether the latent variables continue to encode meaningful residual structure, or whether their influence diminishes once the supervised vector $y$ captures the dominant modes of variation. Training the VAE is carried out on the same dataset and under the same hyperparameters as those in Section~\ref{sec:threefeatures}. 

\autoref{fig:error_distribution_LSCT} shows the distribution of absolute generation errors across all four features. The maximum absolute errors are approximately $5.7\times 10^{-4}$ for level, $4.6\times 10^{-3}$ for slope, $9.4\times 10^{-3}$ for curvature, and $2.0\times 10^{-3}$ for term-structure. The majority of feature deviations fall below $10^{-2}$. The generation accuracy is slightly reduced, but remains within acceptable ranges. The generator maintains stable performance even when four features are simultaneously controlled. 

\begin{figure}[htbp]
    \centering
    \includegraphics[width=.8\linewidth]{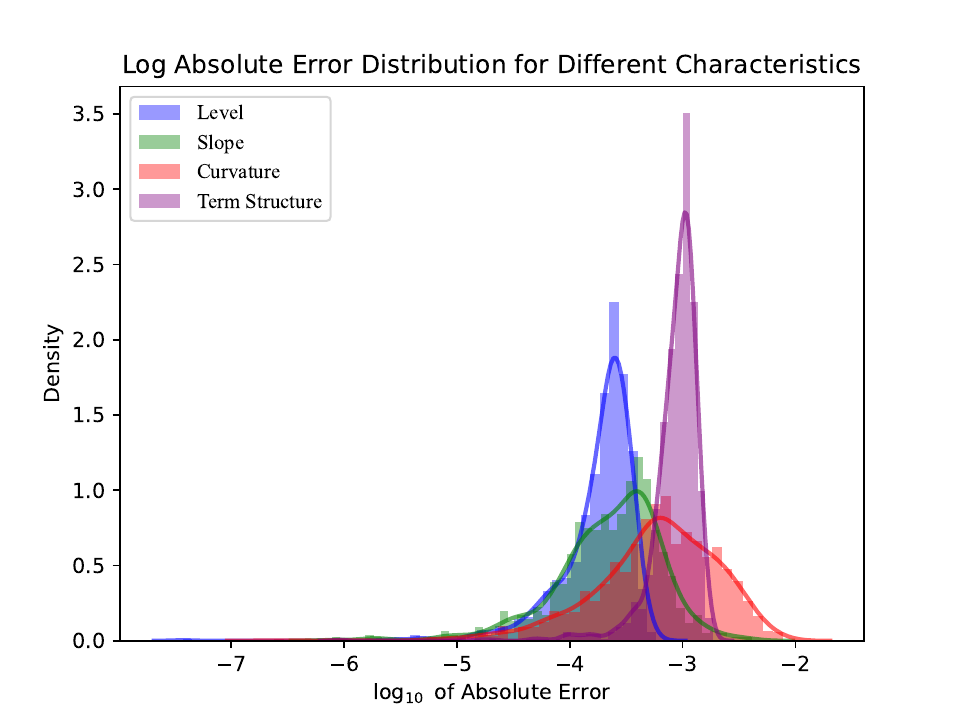}
    \caption{Histogram of log absolute errors for simultaneous control of level, slope, curvature, and term-structure characteristics.}
    \label{fig:error_distribution_LSCT}
\end{figure}

Figures~\ref{fig:2d_level_LSCT},~\ref{fig:2d_slope_LSCT},~\ref{fig:2d_curvature_LSCT} and~\ref{fig:2d_term_structure_LSCT} illustrate how each of the four features can be targeted individually while the others remain fixed. For example, when varying curvature $y_C$ while the other three features are held fixed, the generated surfaces exhibit systematic changes in the smile profile along moneyness. when $y_C$ increases, the implied volatility curve across strikes becomes more convex, with higher volatilities in both deep in- and out-of-the-money regions relative to the at-the-money point. Conversely, decreasing $y_C$ flattens the smile and eventually produces concave shapes, where intermediate strikes show lower volatility than both tails. ~\autoref{fig:2d_curvature_LSCT} shows the above progression that larger positive curvature values produce pronounced U-shaped surfaces, while negative values lead to inverted, concave profiles. These changes occur without material drift in level, slope, or term-structure, confirming that curvature can be manipulated independently. The only notable limitation arises in strongly concave scenarios, where errors grow slightly due to the scarcity of such shapes in the training data.  Similar disentangled behavior is observed when sweeping level, slope, or term-structure in turn. These examples confirm that the four shape features can be manipulated independently through the controllable variables.

\begin{figure}[htbp]
    \centering
    \includegraphics[width=\linewidth]{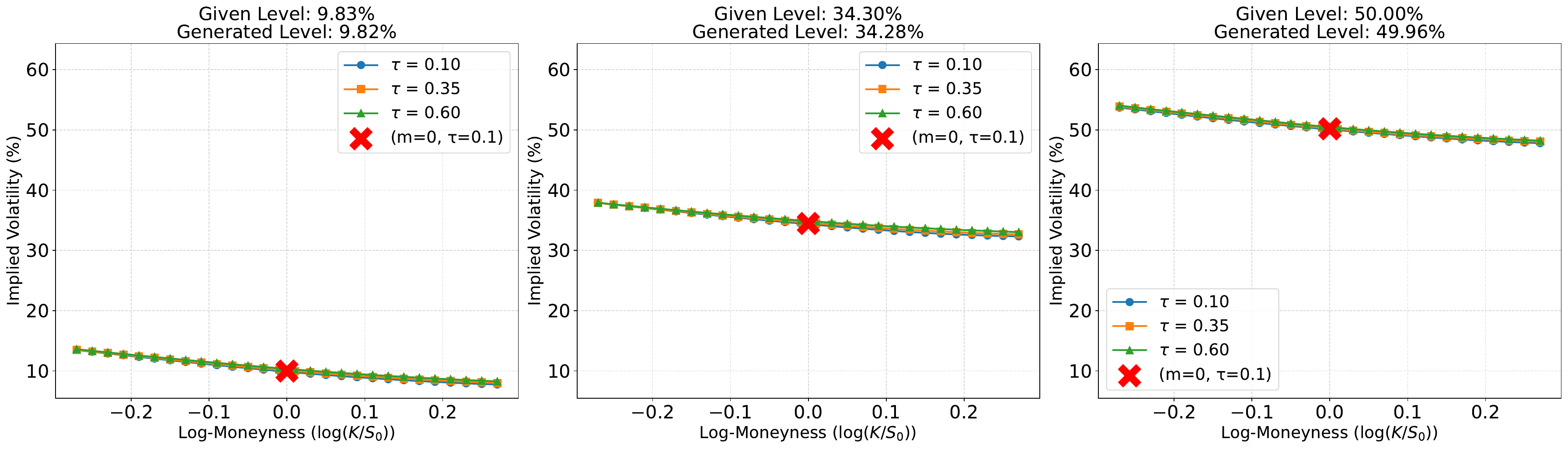}
    \caption{From left to right, given/generated $y_L=9.83/9.82,34.30/34.28,50.00/49.96(\%)$ with fixed $\mathbf{z}$.}
    \label{fig:2d_level_LSCT}
\end{figure}

\begin{figure}[htbp]
    \centering
    \includegraphics[width=\linewidth]{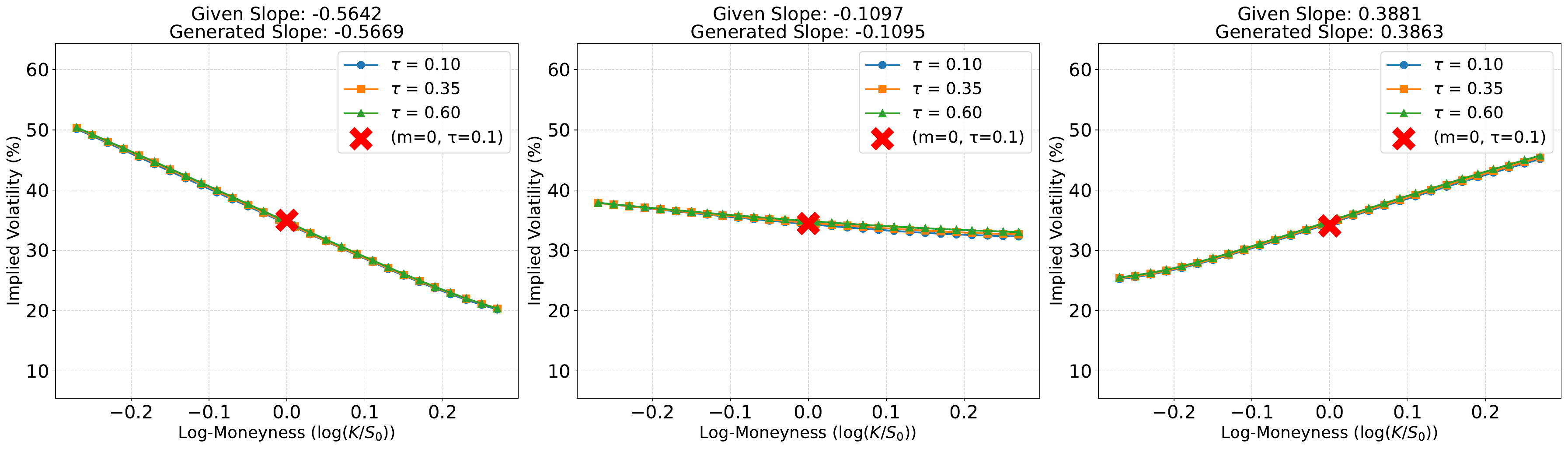}
    \caption{From left to right, given/generated $y_S=-0.5642/-0.5669,-0.1097/-0.1095,0.3881/0.3863$ with fixed $\mathbf{z}$.}
    \label{fig:2d_slope_LSCT}
\end{figure}

\begin{figure}[htbp]
    \centering
    \includegraphics[width=\linewidth]{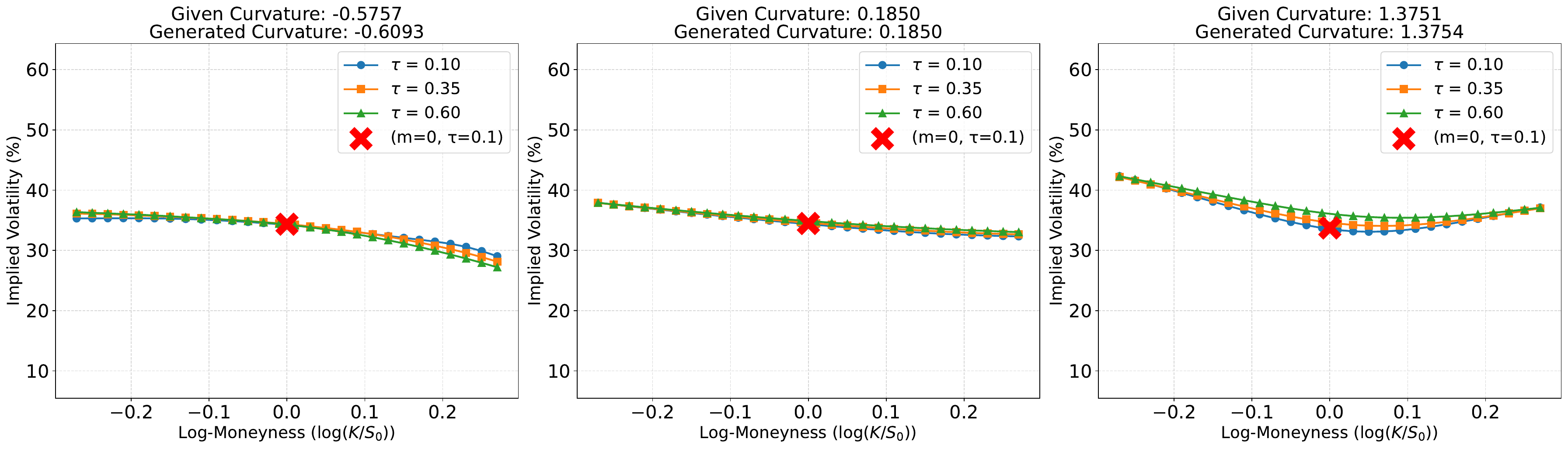}
    \caption{From left to right, given/generated $y_C=-0.5757/-0.6093,0.1850/0.1850,1.3751/1.3754$ with fixed $\mathbf{z}$.}
    \label{fig:2d_curvature_LSCT}
\end{figure}

\begin{figure}[htbp]
    \centering
    \includegraphics[width=\linewidth]{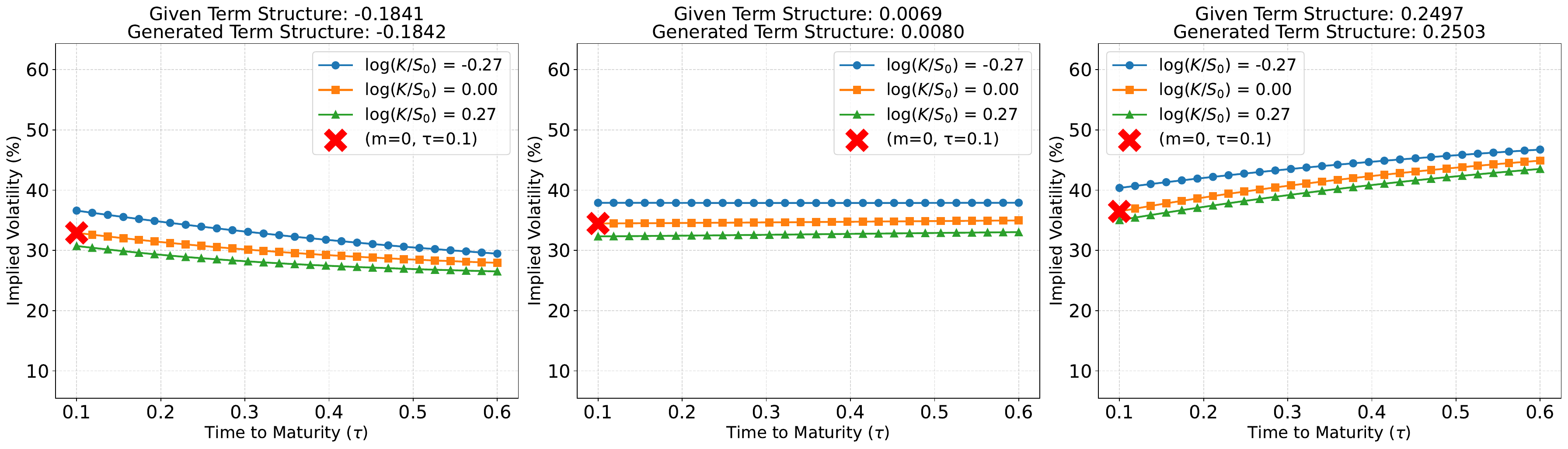}
    \caption{From left to right, given/generated $y_T=-0.1841/-0.1842,0.0069/0.0080,0.2497/0.2503$ with fixed $\mathbf{z}$.}
    \label{fig:2d_term_structure_LSCT}
\end{figure}

At the same time, the latent space $\mathbf{z}$ no longer causes substantial additional shape variations. First, we examine the impact of varying $z_1$ on the generated IVS shape characteristics, while freezing $\mathbf{y}$  and setting the other latent dimensions of $\mathbf{z}$ to be zero. As shown in Table~\ref{tab:z1_variation}, when $z_1$ changes from $-8.00$ to $8.00$ (equivalent to 8 standard deviations), the IVS level  fluctuates by around $2\times10^{-4}$ (between  $0.3874$ and $0.3872$), the slope $1.2\times10^{-3}$, the curvature $1.5\times10^{-3}$, and the term structure $5\times10^{-4}$, indicating negligible variation across all four features. This also demonstrates that $\mathbf{y}$ provides precise control over the specified shape characteristics: once $\mathbf{y}$ is fixed, the latent space $\mathbf{z}$ has little influence on those shape values determined by $\mathbf{y}$. The corresponding IVSs, shown in \autoref{fig:z1_effect_LSCT}, confirm that varying $z_1$ from $-8$ to $8$ does not yield  significant shape changes to the generated IVSs. Similar behaviour is observed when varying the rest latent components $(z_2,z_3,z_4,z_5)$.

\begin{table}[htbp]
\centering
\caption{Effect of varying $z_1$ on IVS shape characteristics while keeping $\mathbf{y}$ fixed and the remaining four  latent dimensions fixed to be zero.}
\label{tab:z1_variation}
\begin{tabular}{cccccc}
\toprule
$z_1$ & Level & Slope & Curvature & Term Structure \\
\midrule
$-8.00$ & $0.3874$ & $-0.2145$ & $0.1353$ & $-0.0229$ \\
$-4.00$ & $0.3874$ & $-0.2142$ & $0.1358$ & $-0.0228$ \\
$0.00$  & $0.3873$ & $-0.2139$ & $0.1362$ & $-0.0226$ \\
$4.00$  & $0.3873$ & $-0.2136$ & $0.1365$ & $-0.0225$ \\
$8.00$  & $0.3872$ & $-0.2134$ & $0.1368$ & $-0.0224$ \\
\bottomrule
\end{tabular}
\end{table}

To summarize, the multiple-feature experiment show that the proposed model can simultaneously control all major stylized factors of IVSs with high precision. Meanwhile,  the  role of the latent variables diminishes: once  four shape characteristics are supervised, the latent space $\mathbf{z}$ ceases to encode interpretable variations in IVS shape, as the controllable vector $y$ accounts for nearly all structured variability present in the training dataset.

\subsection{Experiment III: Arbitrage-Free Conditions}
\label{sec:exp3}

The previous experiments (Sections~\ref{sec:exp1} and \ref{sec:exp2}) demonstrated that our controllable VAE framework can precisely adjust one or more shape features of the IVS. While these results establish controllability and interpretability, financial validity requires an additional criterion: generated IVSs must also satisfy the classical no-arbitrage conditions of option pricing. In practice, even a surface that matches desired feature values is unusable if it admits static arbitrage opportunities. Since the VAE training objective does not enforce these constraints directly, arbitrage violations may still arise,

To investigate this, we generate 60,000 IVSs  by decoding pairs $(\mathbf{y}, \mathbf{z})$ under the three-feature model from Section~\ref{sec:threefeatures}.  The controllable variables $\mathbf{y}=({y_L,y_S,y_T})$ are randomly sampled within an extended hyper-rectangular region based on the training dataset, and $\mathbf{z}$ is drawn either from the central region of the Gaussian prior (within three standard deviations) or from its tails.  Arbitrage violations are most likely to occur when the latent variables $\mathbf{z}$ are sampled from the tails of the Gaussian prior, i.e., far from the training manifold. Under in-distribution sampling of the controllable variables, no arbitrage violations are observed. Specifically, when $\mathbf{y}$ lies within the convex hull of the dataset label space and $\mathbf{z}$ is drawn from the central region of $\mathcal{N}(0,1)$, all 60,000 generated IVSs satisfy both the calendar-spread and butterfly conditions. In \autoref{fig:convexhull}, the orange region denotes the dataset label distribution, while the blue region shows its convex hull.\footnote{The convex hull of a set is the smallest convex set containing it; see \href{https://en.wikipedia.org/wiki/Convex_hull}{Convex Hull}.}

\begin{figure}[htbp]
    \centering
    \includegraphics[width=0.8\linewidth]{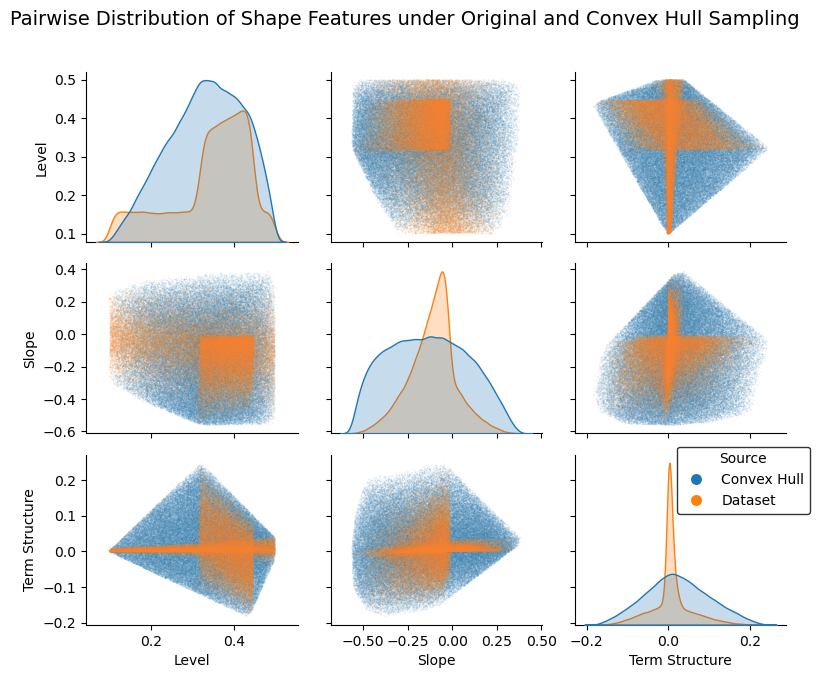}
    \caption{Visualization of the dataset feature domain (orange) and its convex hull (blue).}
    \label{fig:convexhull}
\end{figure}

When sampling more broadly, violations begin to emerge. When $\mathbf{y}$ is drawn  from the extended hyper-rectangular domain while $\mathbf{z}$ is sampled from the full Gaussian prior, approximately $9\%$ of the generated IVSs ($5,412$ out of $60,000$) contains data points which violate either the calendar-spread condition, the butterfly condition, or both. These violations predominantly arise when $\mathbf{z}$ takes extreme tail values. Importantly, however, the  majority (about $91\%$) of IVSs remain arbitrage-free even under these out-of-distribution settings. 

For the $5,412$ IVSs that initially violated arbitrage conditions, we applied the post-generation correction algorithm described in Section~\ref{sec:repair_method}. The correction procedure successfully repaired $2,462$ cases ($45.5\%$), see \autoref{tab:arb_repair}. The remaining $2,950$ surfaces could not be corrected, often due to severe distortions in the tails that are inconsistent with the training manifold.

\begin{table}[htbp]
\centering
\caption{Summary of arbitrage violations and correction performance on 60,000 generated IVSs.}
\label{tab:arb_repair}
\begin{tabular}{lccc|c}
\toprule
 & Valid & Violations& Violations after repair  & Repaired violations\\
\midrule
Count & 54,588 & 5,412  & 2,950 & 2,462 \\
Percentage & $91.0\%$ &  $9.0\%$  & 4.9\% &  45.5\% (Correction rate)\\
\bottomrule
\end{tabular}
\end{table}

To conclude,  the controllable VAE framework produces desired  features while generating implied volatility surfaces that are largely arbitrage-free. Even when sampling beyond the dataset range, arbitrage violations remain rare and can be corrected effectively with the post-generation correction method.

\section{Conclusions and Discussions}\label{sec:conclusion}
In this paper, we have presented a controllable generative modelling framework based on a variational autoencoder for generating synthetic implied volatility surfaces  with desired and quantifiable features. Unlike existing generative approaches, which largely focus on data-driven replication of IVSs, our framework introduces explicit and interpretable control over surface shape characteristics, thereby enabling a transparent link between market stylized facts and generated outcomes.

Through extensive simulation experiments, the framework demonstrates several favorable properties. First, the explicit control mechanism allows the model to generate IVSs that match the targeted feature values with high accuracy, and it performs reliably whether controlling a single feature or multiple features simultaneously. Second, the two-component design of comprising the controllable variables and the residual latent variables  enables implied volatility surfaces that satisfy targeted shape features while preserving diversity. The latent space captures residual structure not explicitly supervised by the controllable variables, which is useful for uncovering  new stylized facts. For example,  when fewer than four shape features are explicitly controlled in our experiments, the model can still discover meaningful hidden features.   Third, the framework exhibits strong generalization: when the controllable variables are sampled within the convex hull of the training distribution, all generated IVSs satisfy arbitrage-free conditions. Even when extrapolating the controllable variables to a broader hyper-rectangular domain while keeping the residual latent variables sampled from the central region of the standard Gaussian distribution, more than 90\% of the generated IVSs remain arbitrage-free. For the remaining small proportion that violate no-arbitrage constraints, our post-generation repair algorithm successfully corrects approximately half.

Several promising directions remain for future research. First, the current polynomial regression struggles to capture cases where the IVS slope explodes at the anchor point, and  adapting the basis functions to handle infinite slopes could address this limitation. Second, new IVS shape features may be incorporated, such as local concavity \citep{alexiou2025pricing}.

In summary, the proposed framework represents a step toward practical generative modeling of implied volatility surfaces, combining interpretability, controllability, and consistency with financial constraints.  The above advantages make the method potentially useful for downstream financial applications, including stress testing, option pricing  and risk management (e.g., under hypothetical scenarios), and data-driven market simulators.

\bibliography{ref2}

\appendix
\section{Training dataset}\label{sec:ParaModels}

The dataset used to train the VAE consists of 60,000 IVSs, synthesized from 30,000 Heston and 30,000 SABR samples.  The distribution of features of the IVSs is summarized below (see \autoref{fig:ivs_feature_distributions}):

\begin{itemize}
    \item \textbf{Level:} Implied volatilities range from 0.098 to 0.500, with a mean of 0.343 and a standard deviation of 0.096, indicating that most IVSs exhibit moderate volatility levels.
    
    \item \textbf{Slope:} The slope, representing the skewness of the smile (first derivative with respect to $m$ at-the-money), ranges from $-0.564$ to $0.388$, with a mean of $-0.110$, reflecting a pronounced left skew on average.
    
    \item \textbf{Curvature:} The curvature (second derivative with respect to $m$) ranges from $-0.576$ to $1.375$, with a mean of $0.185$ and a standard deviation of $0.249$. This indicates a wide variety of smile shapes, including both convex and mildly concave profiles.
    
    \item \textbf{Term Structure:} The slope of the term structure (derivative with respect to $\tau$ at-the-money) ranges from $-0.184$ to $0.250$, with a mean of $0.0069$ and a standard deviation of $0.0483$, suggesting that most surfaces exhibit nearly flat short-term term structures.
\end{itemize}
\begin{figure}
    \centering
    \includegraphics[width=0.8\linewidth]{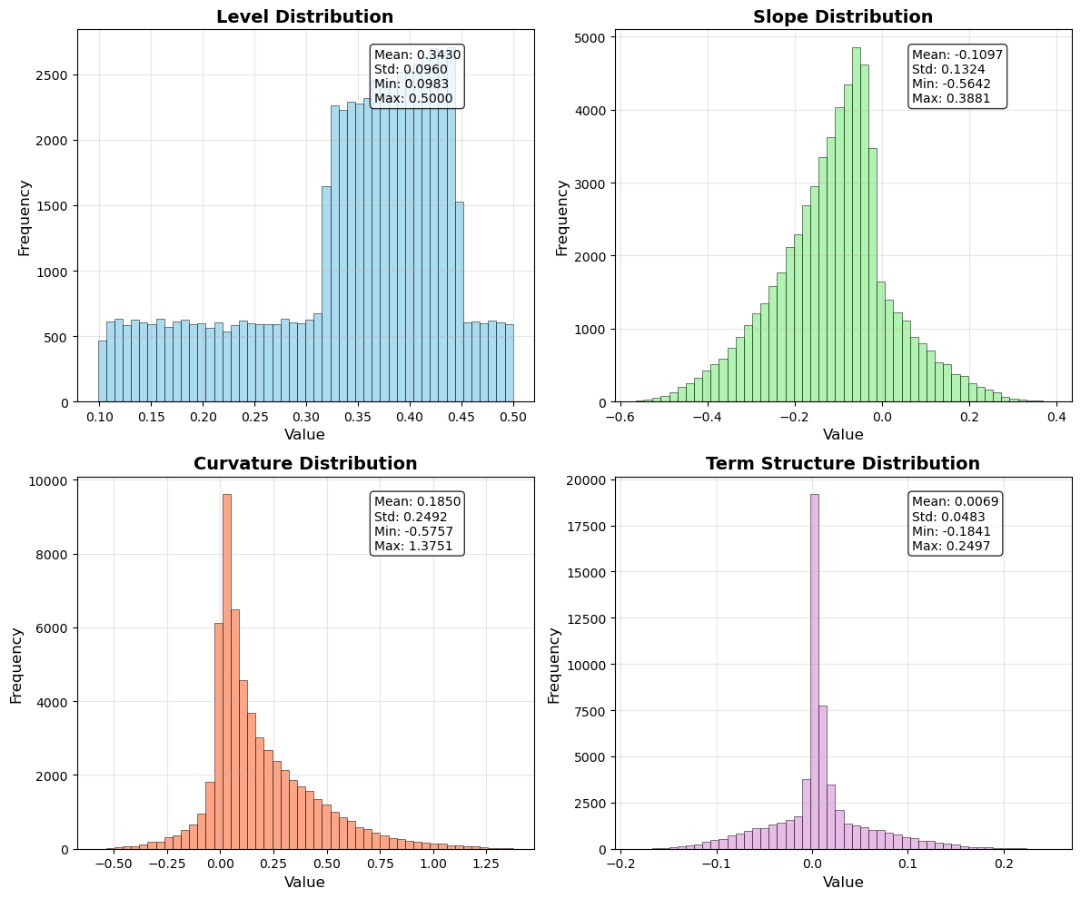}
    \caption{
    Distribution of features extracted from the training IVSs: \textbf{Level} (top left), \textbf{Slope} (top right), \textbf{Curvature} (bottom left), and \textbf{Term Structure} (bottom right). Each subplot shows the empirical histogram along with summary statistics including mean, standard deviation, and range.}
    \label{fig:ivs_feature_distributions}
\end{figure}

\subsection{Heston model}\label{sec:Heston}
In this part, we provide an overview of the Heston model, which can fit market behaviour well. In experiments section (\autoref{sec:Numerical}), we employ the Heston model to simulate a real-world dataset for evaluating our approach.

The Heston model (Equation~\eqref{eq:heston}) is a widely recognized stochastic volatility framework in financial modelling. It assumes that the volatility of the underlying asset is itself stochastic and follows the Cox-Ingersoll-Ross (CIR) process \cite{doi:10.1142/9789812701022_0005}. The model is represented by the following set of stochastic differential equations:

\begin{subequations}\label{eq:heston}
\begin{align}
&\mathrm{d} S(t) = r S(t) \mathrm{~d} t + \sqrt{v(t)} S(t) \, \mathrm{d} W_x^{\mathbb{Q}}(t), \quad S\left(t_0\right) = S_0 > 0, \\ 
&\mathrm{d} v(t) = \kappa(\bar{v} - v(t)) \mathrm{~d} t + \beta_2 \sqrt{v(t)} \, \mathrm{d} W_v^{\mathbb{Q}}(t), \quad v\left(t_0\right) = v_0 > 0, \\ 
&\mathrm{d} W_x^{\mathbb{Q}}(t) \, \mathrm{d} W_v^{\mathbb{Q}}(t) = \rho \, \mathrm{~d} t,
\end{align}
\end{subequations}
where \( S(t) \) represents the price of the underlying asset at time \( t \), and \( v(t) \) denotes the instantaneous variance. The risk-free interest rate \( r \) is assumed to be constant. The parameters \( \kappa \), \( \bar{v} \), and \( \gamma \) describe the mean-reversion rate, the long-term average variance, and the volatility of the variance (commonly known as the volatility of volatility), respectively. The notations \( W_x^{\mathbb{Q}}(t) \) and \( W_v^{\mathbb{Q}}(t) \) are Brownian motions under the risk-neutral probability measure \( \mathbb{Q} \), representing randomness in the asset price and variance dynamics. The correlation coefficient \( \rho \) characterizes the relationship between these two Brownian motions. The initial values of the asset price and variance are \( S_0 \) and \( v_0 \), respectively.

Unlike the Black-Scholes model, the Heston model does not admit closed-form analytical solutions. However, it is often considered semi-analytical due to its tractable mathematical structure. By taking the logarithm of the asset price, the equations in Equation~\eqref{eq:heston} become affine, allowing for the derivation of a characteristic function. This property enables the efficient computation of option prices using numerical techniques such as the COS method \cite{doi:10.1137/080718061}. As a result, the Heston model serves as a tool for capturing complex market dynamics, making it suitable for our experiments.

\autoref{tab:hestonpara} presents the parameter ranges used in our Heston dataset setups. We set $S_0 = 1$ without loss of generality, since our analysis focuses exclusively on the shape characteristics of the IVS. Setting $r = 0$ eliminates the impact of the interest rate, ensuring that $S_t = \mathbb{E}(S_0)$ under the risk-neutral measure. Consequently, $\log\left(\frac{K}{S(t)}\right)$ becomes equivalent to $\log\left(\frac{K}{S_0}\right)$ in the IVS. The remaining parameters are chosen empirically, as market data consistently fall within these ranges. To get our simulated Heston IVS dataset, we first randomly select parameters within the specified range, then numerically compute the corresponding implied volatilities for each $\tau$ and $K$. Finally, a smooth surface is fitted through these points, forming our training dataset.

\begin{table}[htpb]
    \centering
    \caption{Parameters of Heston model}
    \begin{tabular}{cc}
    \toprule
Parameter & Setting        \\
\midrule
$S_0$     & 1              \\
$\log \frac{K}{S_0}$  & [-0.27, 0.27]     \\
$\tau$       & [0.1, 0.6]     \\
$r$       & 0.      \\
$\rho$    & [-0.9, -0.1]   \\
$\bar{v}$ & [0.1, 0.3]     \\
$\kappa$  & [1.0, 2.0]     \\
$\gamma$  & [0.1, 0.9]     \\
\bottomrule
\end{tabular}
\label{tab:hestonpara}
\end{table}

\subsection{SABR model \cite{hagan2002managing}}
SABR is another kind of stochastic volatility model. In the $T$-forward measure $\mathbb{Q}^T$, the SABR model can be described by the following system of stochastic differential equations (SDEs):
\begin{align}
dS_F(t,T) &= \sigma(t)\,\bigl(S_F(t,T)\bigr)^{\beta}\,dW_F(t), 
\quad S_F(t_0,T) = s_F(0),\\
d\sigma(t) &= \gamma\,\sigma(t)\,dW_{\sigma}(t), 
\quad \sigma(t_0) = \alpha,
\end{align}
where the Brownian motions satisfy
\[
dW_F(t)\,dW_{\sigma}(t) \;=\; \rho\,dt.
\]

Typically, $\sigma(t)$ is modeled as a lognormal process. Furthermore, for constant $\sigma(t)$, the forward price $S_F(t,T)$ follows a CEV (constant elasticity of variance) process. Consequently, when conditioning on the paths of $\sigma(t)$ over the interval $0 \leq t \leq T$, the resulting SABR dynamics for $S_F(t,T)$ can also be interpreted as a CEV process. Both \(\beta\) and \(\rho\) influence the shape of the implied volatility skew. In practical applications, \(\beta\) is usually kept constant, while \(\rho\) is determined through calibration. The parameter \(\alpha\) dictates the overall level of the implied volatility smile, whereas \(\gamma\) controls the extent of its curvature \cite{oosterlee2019mathematical}.

As stated on page 112 of~\cite{oosterlee2019mathematical}, the implied volatility under the SABR model can be approximated as
\begin{equation}
\label{eq:sabr_iv}
    \hat{\sigma}(T,K)=\frac{\hat{a}(K)\hat{c}(K)}{g(\hat{c}(K))}
    \times\left[1+\left(\frac{(1 - \beta)^2}{24}\frac{\alpha^2}{(S_F(t_0)K)^{1 - \beta}}+\frac{1}{4}\frac{\rho\beta\gamma\alpha}{(S_F(t_0)K)^{\frac{1 - \beta}{2}}}+\frac{2 - 3\rho^2}{24}\gamma^2\right)T\right],
\end{equation}
where
\begin{equation}
\label{eq:a_hat}
    \hat{a}(K)=\frac{\alpha}{(S_F(t_0)\cdot K)^{\frac{1 - \beta}{2}}\left(1+\frac{(1 - \beta)^2}{24}\log^2\left(\frac{S_F(t_0)}{K}\right)+\frac{(1 - \beta)^4}{1920}\log^4\left(\frac{S_F(t_0)}{K}\right)\right)},
\end{equation}
and
\begin{equation}
\label{eq:c_hat}
    \hat{c}(K)=\frac{\gamma}{\alpha}(S_F(t_0))^{\frac{1 - \beta}{2}}\log\left(\frac{S_F(t_0)}{K}\right), \quad g(x)=\log\left(\frac{\sqrt{1 - 2\rho x+x^2}+x - \rho}{1 - \rho}\right).
\end{equation}

In the special case of at-the-money (ATM) options, i.e., when \(S_F(t_0) = K\), the approximation simplifies to
\begin{equation}
\label{eq:sabr_iv_atm}
    \hat{\sigma}(T,K)\approx\frac{\alpha}{(S_F(t_0))^{1 - \beta}}\left(1+\left[\frac{(1 - \beta)^2}{24}\frac{\alpha^2}{(S_F(t_0))^{2 - 2\beta}} + \frac{1}{4}\frac{\rho\beta\alpha\gamma}{(S_F(t_0))^{1 - \beta}} + \frac{2 - 3\rho^2}{24}\gamma^2\right]T\right).
\end{equation}

To construct the SABR IVS dataset, we randomly sample parameters from the ranges shown in \autoref{tab:sabrpara}, and then compute the implied volatilities using the expressions in Equations~\eqref{eq:sabr_iv}–\eqref{eq:sabr_iv_atm}. This procedure yields a synthetic SABR-based IVS dataset for subsequent training and evaluation.
\begin{table}[htpb]
    \centering
    \caption{Parameters of SABR model}
    \begin{tabular}{cc}
    \toprule
Parameter & Setting        \\
\midrule
$S_0$     & 1              \\
$\log \frac{K}{S_0}$  & [-0.27, 0.27]     \\
$\tau$       & [0.1, 0.6]     \\
$r$       & 0              \\
$\beta$   & [0.1, 1.0]     \\
$\alpha$  & [0.1, 0.5]     \\
$\rho$    & [-0.9, 0.9]    \\
$\gamma$  & [0.1, 0.9]     \\
\bottomrule
\end{tabular}
\label{tab:sabrpara}
\end{table}

\section{Bivariate Taylor expansion}
\label{sec:anchor}

As introduced in \autoref{sec:ShapeFeatures}, the \textit{anchor point} defined by \( \tau \rightarrow 0^+ \) and \( m = 0 \) contains rich structural information and serves as a representative location on the IVS. This point has been extensively studied (e.g., \cite{ait2021implied, alos2021malliavin}) and plays a critical role in reconstructing the full IVS, distinguishing among different option pricing models, and capturing key aspects of market dynamics.

In this section, we provide an approach for numerically approximating and quantifying shape characteristics at the anchor point via Taylor expansion and polynomial regression. Recall that, for non-negative integers \( i \) and \( j \), we define the shape characteristics \( \Sigma_{i,j} \) as
\begin{equation}
    \Sigma_{i,j} = \lim_{\tau \rightarrow 0^+} \frac{\partial^{i+j} g}{\partial \tau^i\, \partial m^j}(\tau, 0),
\end{equation}
where \( \tau \) is the time to maturity, \( m = \log(K/S_0) \) is the log-moneyness, and the definition of $g$ is the same as Equation~\eqref{eq:g}. In our implementation, we focus on four fundamental shape characteristics: the level \( \Sigma_{0,0} \), skewness \( \Sigma_{0,1} \), curvature \( \Sigma_{0,2} \), and term-structure slope \( \Sigma_{1,0} \). These are collectively denoted as
\[
\mathbf{G} \subseteq [\Sigma_{0,0}, \Sigma_{0,1}, \Sigma_{0,2}, \Sigma_{1,0}].
\]

The shape characteristics around this anchor point can be estimated using the methodology outlined in Section 2 of \cite{ait2021implied}, which proceeds as follows:

\begin{enumerate}
    \item We first express the IVS locally using a bivariate Taylor expansion around the anchor point:
    \begin{equation}\label{eq:Taylor}
        \mathbf{x}^{(J, \mathbf{L}(J))}(\tau, m) = \sum_{j=0}^J \sum_{i=0}^{L_j} \beta^{(i, j)} \tau^i m^j, \quad \text{where} \quad \beta^{(i, j)} = \frac{\Sigma_{i,j}}{i!\,j!}.
    \end{equation}
    Here, \( J \in \mathbb{N} \) and \( \mathbf{L}(J) = (L_0, L_1, \dots, L_J) \) are finite expansion orders in each direction.
    
    \item Given a dataset consisting of \( n \) daily IVSs, we fit the expansion in Equation~\eqref{eq:Taylor} via polynomial regression on each day \( l = 1, \dots, n \). The observed IVS values \( \mathbf{x}^{\mathrm{data}}(\tau_l^{(s)}, m_l^{(s)}) \) are approximated as:
    \begin{equation}\label{eq:regr}
        \mathbf{x}^{\mathrm{data}}(\tau_l^{(s)}, m_l^{(s)}) = \sum_{j=0}^J \sum_{i=0}^{L_j} \beta_l^{(i, j)} (\tau_l^{(s)})^i (m_l^{(s)})^j + \epsilon_l^{(s)},
    \end{equation}
    where \( s = 1, \dots, n_l \) indexes the IVS grid points on day \( l \), and \( \epsilon_l^{(s)} \) denotes zero-mean i.i.d. observation noise.
    
    \item Finally, the estimated shape characteristics on day \( l \) are recovered as:
    \begin{equation}\label{eq:labley}
        [\Sigma_{i,j}]_l^{\mathrm{data}} = i!\,j!\,\hat{\beta}_l^{(i,j)}, \quad \text{for } i,j \geq 0.
    \end{equation}
\end{enumerate}

\section{Derivation of conditional VAE}
\label{sec:GaussianLoss}

A central task in training generative models is to estimate the model parameters $\theta$ such that the resulting distribution $p_\theta(\mathbf{x})$ closely approximates the true data-generating distribution $p^*(\mathbf{x})$. Among several available inference strategies, a common and intuitive approach is to maximize the log-likelihood over a dataset of $n$ i.i.d. samples $\{\mathbf{x}_i\}_{i=1}^{n}$. The optimal parameters $\theta^*$ are obtained by solving:

\begin{equation}
\theta^* = \arg\max_{\theta \in \Theta} \sum_{i=1}^{n} \log p_\theta(\mathbf{x}_i).
\end{equation}

This maximum likelihood estimation (MLE) method is widely used due to its simplicity and strong statistical guarantees. However, it becomes computationally intractable when $p_\theta(\mathbf{x})$ includes latent variables, as evaluating the marginal likelihood requires integrating over those latent variables.

In models with latent variables, the marginal likelihood is computed by integrating out both the controllable variables $\mathbf{y}$ and the latent variables $\mathbf{z}$:
\begin{equation}
\log p_\theta(\mathbf{x}) = \log \iint p_\theta(\mathbf{x}, \mathbf{y}, \mathbf{z}) \, d\mathbf{y} \, d\mathbf{z} = \log \iint p(\mathbf{z}) \cdot p(\mathbf{y}) \cdot p_\theta(\mathbf{x} \mid \mathbf{y}, \mathbf{z}) \, d\mathbf{y} \, d\mathbf{z}.
\end{equation}
While this formulation is exact and theoretically sound, evaluating such high-dimensional integrals is analytically and numerically intractable in most practical settings.

To overcome this limitation, variational inference \cite{blei_variational_2017, kingma2014semi} introduces a tractable, parameterized variational distribution $q_\phi(\mathbf{z} \mid \mathbf{x}, \mathbf{y})$ to approximate the true posterior $p(\mathbf{z} \mid \mathbf{x}, \mathbf{y})$. Instead of directly estimating $\log p_\theta(\mathbf{x})$, we now consider the joint distribution $\log p_\theta(\mathbf{x}, \mathbf{y})$, where $\mathbf{y}$ is assumed to be observed in our experimental setting. This shift allows us to condition on $\mathbf{y}$ explicitly and derive a tighter variational bound.

The log-likelihood of the joint distribution can be expressed using the variational distribution as:
\begin{equation}\label{eq:pxy}
\log p_\theta(\mathbf{x}, \mathbf{y}) = \mathbb{E}_{\mathbf{z} \sim q_{\phi}(\mathbf{z} \mid \mathbf{x}, \mathbf{y})} \left[ 
\frac{\log p_\theta(\mathbf{x}, \mathbf{y}, \mathbf{z})}{q_{\phi}(\mathbf{z} \mid \mathbf{x}, \mathbf{y})} 
\cdot 
\frac{q_{\phi}(\mathbf{z} \mid \mathbf{x}, \mathbf{y})}{p(\mathbf{z} \mid \mathbf{x}, \mathbf{y})}
\right],
\end{equation}
where $\theta$ and $\phi$ denote the parameters of the decoder and encoder networks, respectively.

By further decomposing Equation~\eqref{eq:pxy}, and using the factorization $p_\theta(\mathbf{x}, \mathbf{y}, \mathbf{z}) = p_\theta(\mathbf{x} \mid \mathbf{z}, \mathbf{y}) \cdot p(\mathbf{z}) \cdot p(\mathbf{y})$ (Equation~\eqref{eq:pxyz}), we obtain:
\begin{equation}\label{eq:p(x, y)}
\begin{aligned}
\log p_\theta(\mathbf{x}, \mathbf{y}) 
=& \underbrace{\mathbb{E}_{\mathbf{z} \sim q_{\phi}(\mathbf{z} \mid \mathbf{x}, \mathbf{y})} \left[
\log p_\theta(\mathbf{x} \mid \mathbf{z}, \mathbf{y}) + \log p(\mathbf{z}) + \log p(\mathbf{y}) - \log q_{\phi}(\mathbf{z} \mid \mathbf{x}, \mathbf{y})
\right]}_{\text{ELBO}} \\
&+ \underbrace{\mathbb{E}_{\mathbf{z} \sim q_{\phi}(\mathbf{z} \mid \mathbf{x}, \mathbf{y})} \left[
\log \frac{q_{\phi}(\mathbf{z} \mid \mathbf{x}, \mathbf{y})}{p(\mathbf{z} \mid \mathbf{x}, \mathbf{y})}
\right]}_{=
\mathrm{KL}\left( q_\phi(\mathbf{z} \mid \mathbf{x}, \mathbf{y}) \parallel p(\mathbf{z} \mid \mathbf{x}, \mathbf{y}) \right)
}.
\end{aligned}
\end{equation}
The first term in Equation~\eqref{eq:p(x, y)} is commonly referred to as the \textbf{E}vidence \textbf{L}ower \textbf{BO}und (ELBO). The second term of Equation~\eqref{eq:p(x, y)} can be also written as 
$
\mathrm{KL}\left( q_\phi(\mathbf{z} \mid \mathbf{x}, \mathbf{y}) \parallel p(\mathbf{z} \mid \mathbf{x}, \mathbf{y}) \right),
$
representing the Kullback--Leibler (KL) divergence between the approximate distribution \( q_\phi(\mathbf{z} \mid \mathbf{x}, \mathbf{y}) \) and the true posterior distribution \( p(\mathbf{z} \mid \mathbf{x}, \mathbf{y}) \). Since the KL divergence is always non-negative and the true posterior is generally intractable, maximizing $\log p_\theta(\mathbf{x}, \mathbf{y})$ is equivalent to maximizing the ELBO.

Therefore, we define the following objective to be minimized:
\begin{equation}\label{eq:ELBO}
    \begin{aligned}
        &\mathcal{L}(\mathbf{x}, \mathbf{y}) \\
        =& - \mathbb{E}_{\mathbf{z} \sim q_\phi(\mathbf{z} | \mathbf{x}, \mathbf{y})}\left[\log p{_\theta}(\mathbf{x} | \mathbf{z}, \mathbf{y}) {- \log q_\phi(\mathbf{z} | \mathbf{x}, \mathbf{y}) + \log p(\mathbf{z})} \right] \\
        =& \underbrace{- \mathbb{E}_{\mathbf{z} \sim q_\phi(\mathbf{z} | \mathbf{x}, \mathbf{y})}\left[\log p{_\theta}(\mathbf{x} | \mathbf{z}, \mathbf{y})\right] }_{\text{Reconstruction Error}}
        \underbrace{{+ KL\left(q_\phi(\mathbf{z} | \mathbf{x}, \mathbf{y}) \parallel p(\mathbf{z})\right)}}_{\text{KL divergence}}.
    \end{aligned}
\end{equation}
where the first term corresponds to the negative expected log-likelihood (reconstruction loss), and the second term regularizes the posterior by encouraging it to remain close to the prior.

To allow more flexible control over the trade-off between reconstruction accuracy and latent space regularization, we introduce a hyperparameter $\beta$ to scale the KL term. This leads to the final form of the loss:

\begin{equation}\label{eq:ELBO_beta}
\mathcal{L}(\mathbf{x}, \mathbf{y}) 
= - \mathbb{E}_{\mathbf{z} \sim q_\phi(\mathbf{z} \mid \mathbf{x}, \mathbf{y})} \left[
\log p_\theta(\mathbf{x} \mid \mathbf{z}, \mathbf{y})
\right] + \beta \cdot \mathrm{KL}\left(q_\phi(\mathbf{z} \mid \mathbf{x}, \mathbf{y}) \parallel p(\mathbf{z}) \right).
\end{equation}

Here, $\phi$ and $\theta$ represent the parameters of the encoder and decoder networks, respectively. The first term measures how well the generated IVS matches the input IVS (reconstruction quality). Under assumptions we later introduce regarding the distributions of $\mathbf{y}$ and $\mathbf{z}$, this term simplifies to the mean squared error (MSE), as detailed in Section~\ref{sec:GaussianLoss}. The second term acts as a regularizer by promoting a more structured latent space. The effect of the hyperparameter $\beta$ on model performance is further investigated in remark of \autoref{para:beta}.

Finally, training is performed by minimizing the overall loss with respect to both encoder and decoder parameters:

\begin{equation}
\arg\min_{\theta, \phi} \mathcal{L}(\mathbf{x}, \mathbf{y}).
\end{equation}

To derive a closed-form, computationally tractable version of the loss function, we assume that both the prior and the variational posterior follow multivariate Gaussian distributions, as defined in Equation~\eqref{eq:prior} and Equation~\eqref{eq:posterior}. As we know that the Gaussian probability density function:

\begin{equation}
p(\boldsymbol{\xi}) = \frac{1}{(2\pi)^{n/2}|\Sigma|^{1/2}} \exp\left\{ -\frac{1}{2} (\boldsymbol{\xi} - \boldsymbol{\mu})^T \Sigma^{-1} (\boldsymbol{\xi} - \boldsymbol{\mu}) \right\},
\end{equation}
where $\boldsymbol{\xi} \in \mathbb{R}^n$ denotes a generic continuous random vector (e.g., $\mathbf{x}$), $\boldsymbol{\mu} \in \mathbb{R}^n$ is the mean, and $\Sigma \in \mathbb{R}^{n \times n}$ is the covariance matrix. 

We now substitute this expression into the reconstruction term of the negative ELBO, i.e., the expected log-likelihood term $- \mathbb{E}_{\mathbf{z} \sim q_\phi(\mathbf{z} \mid \mathbf{x}, \mathbf{y})} \log p_\theta(\mathbf{x} \mid \mathbf{z}, \mathbf{y})$ as defined in Equation~\eqref{eq:Lxy}. Assuming a unit covariance $\Sigma = \mathbf{I}$, the expression simplifies as follows:

\begin{equation}\label{eq:logpxz}
\begin{aligned}
    &- \mathbb{E}_{\mathbf{z} \sim q_\phi(\mathbf{z} \mid \mathbf{x}, \mathbf{y})} \log p_\theta(\mathbf{x} \mid \mathbf{z}, \mathbf{y}) \\
    &= \frac{1}{L} \sum_{l=1}^{L} \left[ -\log p_\theta(\mathbf{x} \mid \mathbf{z}^{(l)}, \mathbf{y}) \right] \\
    &= -\log \frac{1}{(2\pi)^{n/2}|\mathbf{I}|^{1/2}} 
    + \frac{1}{2L} \sum_{l=1}^{L} (\mathbf{x} - \boldsymbol{\mu}_{z})^T (\mathbf{x} - \boldsymbol{\mu}_{z}) \\
    &= \text{const} + \frac{1}{2L} \sum_{l=1}^{L} \left\| \mathbf{x} - \hat{\mathbf{x}}^{(l)} \right\|_2^2,
\end{aligned}
\end{equation}

where $\hat{\mathbf{x}}^{(l)} = \boldsymbol{\mu}_{z} = \boldsymbol{\mu}_{f_\theta(\mathbf{y}, \mathbf{z}^{(l)})}$ denotes the $l$-th output samples decoded by the $l$-th $\mathbf{z}$. In practice, it is common to use $L = 1$, as empirical results have shown this to be sufficient for stable training~\cite{kingma2013auto}. Thus, the reconstruction term reduces to the standard mean squared error (MSE) between the input $\mathbf{x}$ and its reconstruction $\hat{\mathbf{x}}$:

\begin{equation}
    \mathcal{L}_{\text{rec}}(\mathbf{x}, \hat{\mathbf{x}}) = \frac{1}{2} \left\| \mathbf{x} - \hat{\mathbf{x}} \right\|_2^2 + \text{const}.
\end{equation}

The second term of the ELBO corresponds to the KL divergence between the variational posterior and the standard Gaussian prior for $\mathbf{z}$:
\begin{equation}
\mathrm{KL} \left( q_\phi(\mathbf{z} \mid \mathbf{x}, \mathbf{y}) \parallel p(\mathbf{z}) \right) = \frac{1}{2} \sum_{j=1}^{d_z} \left( \mathbf{\mu}_{z_j}^2 + \mathbf{\sigma}_{z_j}^2 - 1 - \log \mathbf{\sigma}_{z_j}^2 \right),
\end{equation}
where \(d_z\) denotes the dimension of \(\mathbf{z}\), and \((\mathbf{\mu}_{z_j}, \mathbf{\sigma}_{z_j}^2)\) represent the mean and diagonal entries of the variance matrix of the posterior distribution over \(\mathbf{z}\), which are also the outputs of the encoder network.

Therefore, the overall training objective--i.e., the negative ELBO--can be simplified as

\begin{equation}\label{eq:final_loss}
\mathcal{L}(\mathbf{x}) = \frac{1}{2}
\left\| \mathbf{x} - \hat{\mathbf{x}} \right\|_2^2
+ \beta \cdot \sum_{j=1}^{d_z} \left( \mathbf{\mu}_{z_j}^2 + \mathbf{\sigma}_{z_j}^2 - 1 - \log \mathbf{\sigma}_{z_j}^2 \right).
\end{equation}

\begin{algorithm}[H]
\caption{Controllable VAE Training Procedure}
\label{algo:DVAE}
\KwIn{
    Labeled dataset $\mathcal{D}_l = \{(\mathbf{x}_i, \mathbf{y}_i)\}_{i=1}^{N_l}$ ($\mathbf{y}_i \in \mathbb{R}^d$), \\
    Latent dimension $d_z$, Batch size $B$, Epochs $T$, \\
    Learning rate $\eta$, KL weight $\beta$;
}
\KwOut{
    Encoder parameters $\phi$;\\
    Decoder parameters $\theta$;
}

\textbf{Initialize:}\\
    Encoder network $q_{\phi}(\mathbf{z}^{(post)}|\mathbf{x, y}) = \mathcal{N}(\mu_{\phi}(\mathbf{x, y}), \text{diag}(\sigma_{\phi}^2(\mathbf{x, y})))$;\\
    Decoder network $p_{\theta}(\mathbf{x}|\mathbf{z, y})$ parameterized by $\theta$;\\
    Prior $p(\mathbf{z}) = \mathcal{N}(\mathbf{0, I})$;\\

\For{$t = 1$ \textbf{to} $T$}{
    Sample batch $\mathcal{B}_l = \{(\mathbf{x}_i, \mathbf{y}_i)\}_{i=1}^B$ from $\mathcal{D}_l$;\\
    
    $\mathcal{L}_{total} \leftarrow +\infty$;\\
    \ForEach{$(\mathbf{x}_i, \mathbf{y}_i) \in \mathcal{B}_l$}{
        $(\mathbf{\mu}, \log\mathbf{\sigma}_z^2) \leftarrow q_{\phi}(\mathbf{x}_i, \mathbf{y}_i)$;\\
        Sample $\mathbf{\epsilon} \sim \mathcal{N}(\mathbf{0, I})$;\\
        $\mathbf{\sigma} \leftarrow \exp\left(\frac{1}{2} \log \mathbf{\sigma}^2\right)$;\\
        $\mathbf{z}^{(post)} \leftarrow \mathbf{\mu} + \mathbf{\sigma} \odot \mathbf{\epsilon}$;\\
        $\hat{\mathbf{x}} \leftarrow \mathbf{\mu}_{\theta}(\mathbf{z}^{(post)}, \mathbf{y}_i)$;\\
        
        $\mathcal{L}_{rec} \leftarrow -\log p_{\theta}(\hat{\mathbf{x}}|\mathbf{z}^{(post)},\mathbf{y}_i) = \text{MSE}(\mathbf{x}, \hat{\mathbf{x}})=\frac{1}{2} \| \mathbf{x} - \hat{\mathbf{x}} \|_2^2$\;
        
        $\mathcal{L}_{KL} \leftarrow \text{KL}(q_{\phi}(\mathbf{z}^{(post)}|\mathbf{x, y}) || p(\mathbf{z}^{prior}))=\frac{1}{2} \sum_{j=1}^{d_z} \left(\mathbf{\sigma}_{z_j}^2 + \mathbf{\mu}_{z_j}^2 - 1 - \log\mathbf{\sigma}_{z_j}^2 \right)$\;
        
        $\mathcal{L}_{total} \leftarrow \mathcal{L}_{total} + \mathcal{L}_{rec} + \beta \cdot \mathcal{L}_{KL}$\;
    }
    
    $\theta \leftarrow \theta - \eta \cdot \nabla_{\theta} \mathcal{L}_{total}$;\\
    $\phi \leftarrow \phi - \eta \cdot \nabla_{\phi} \mathcal{L}_{total}$;
}
\end{algorithm}

\begin{algorithm}[H]
\caption{Latent Space Optimization for Arbitrage-Free Correction (L-BFGS)}
\label{algo:ArbFree}
\KwIn{
    A generated IVS $\mathbf{x}_i$;  
    Corresponding control variable $\mathbf{y}_i$;  
    Latent variable $ \{\mathbf{z}_i\}$;
    Decoder network $\text{Decoder}(\cdot)$;  
    Maximum iterations $T$;  
    Learning rate $\eta$;  
    Gradient tolerance $\epsilon_1$;  
    Loss change tolerance $\epsilon_2$.
}
\KwOut{
    Arbitrage-free repaired IVSs $\mathbf{X}_{\text{repaired}}$ and optimized latent vectors $\mathbf{Z}_{\text{optimized}}$.
}

\For {$i = 1$ \textbf{to} $N$}{
    \If{$\mathbf{x}_i$ satisfies both calendar spread and butterfly conditions}{
        Continue to next IVS\;
    }
    \Else{
        $\mathcal{L}_{\text{total}}^{(0)} \leftarrow +\infty$\;
        \For{$t = 1$ \textbf{to} $T$}{
            Compute total loss:
            \[
            \mathcal{L}_{\text{total}}^{(t)} = \mathcal{L}_{\text{Monotonicity}} + \mathcal{L}_{\text{Butterfly}} + \mathcal{L}_{\text{MSE}}.
            \]
            \If{$\|\nabla \mathcal{L}_{\text{total}}\| < \epsilon_1$ \textbf{or} $|\mathcal{L}_{\text{total}}^{(t)} - \mathcal{L}_{\text{total}}^{(t-1)}| < \epsilon_2$}{
                \textbf{break}\;
            }
            Update latent vector:
            \[
            \mathbf{z}_i \leftarrow \mathbf{z}_i - \eta \cdot \nabla_{\mathbf{z}_i} \mathcal{L}_{\text{total}}.
            \]
            Decode:
            \[
            \mathbf{x}_i \leftarrow \text{Decoder}(\mathbf{z}_i, \mathbf{y}_i).
            \]
        }
    }
}
\end{algorithm}

\section{Additional plots in numerical results}
\label{sec:add_res}

\subsection{Experiment I}
\begin{figure}[htbp]
    \centering
    \includegraphics[width=\linewidth]{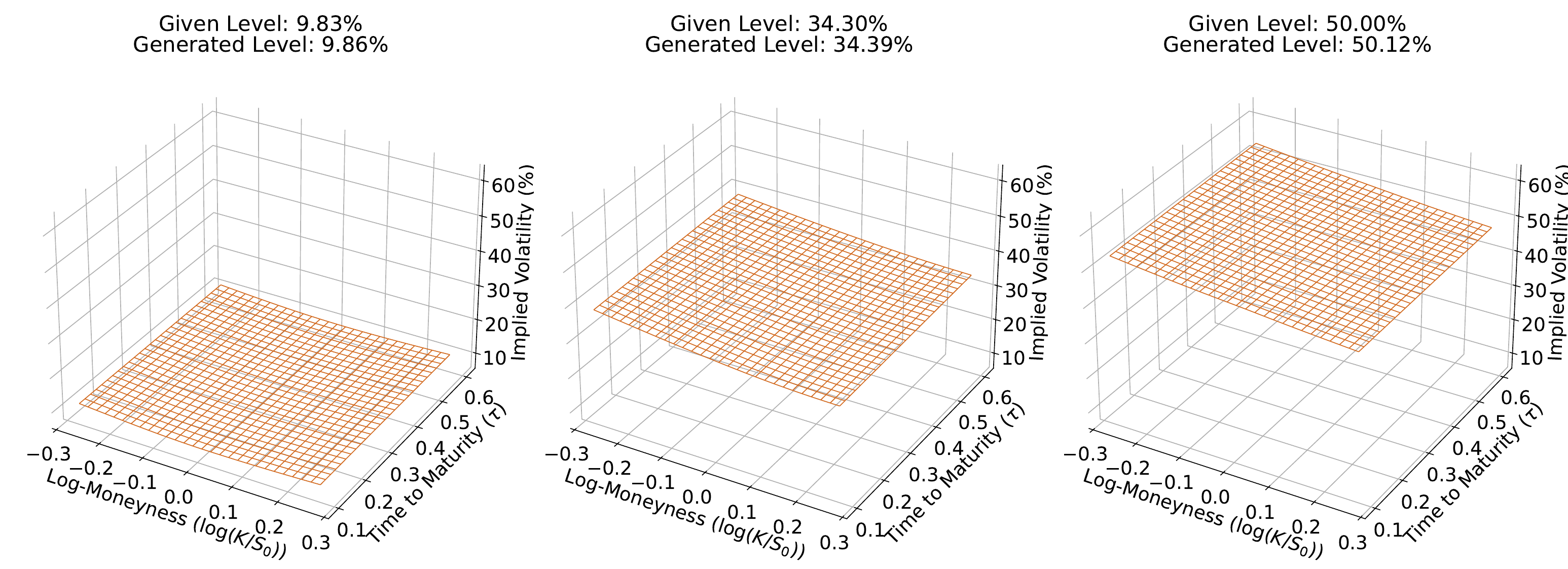}
    \caption{3D IVS varying \textit{level} feature ($y_L$) while latent variables $\mathbf{z}$ fixed.}
    \label{fig:3d_Level_HS_min_mean_max}
\end{figure}

\begin{figure}[htbp]
    \centering
    \includegraphics[width=\linewidth]{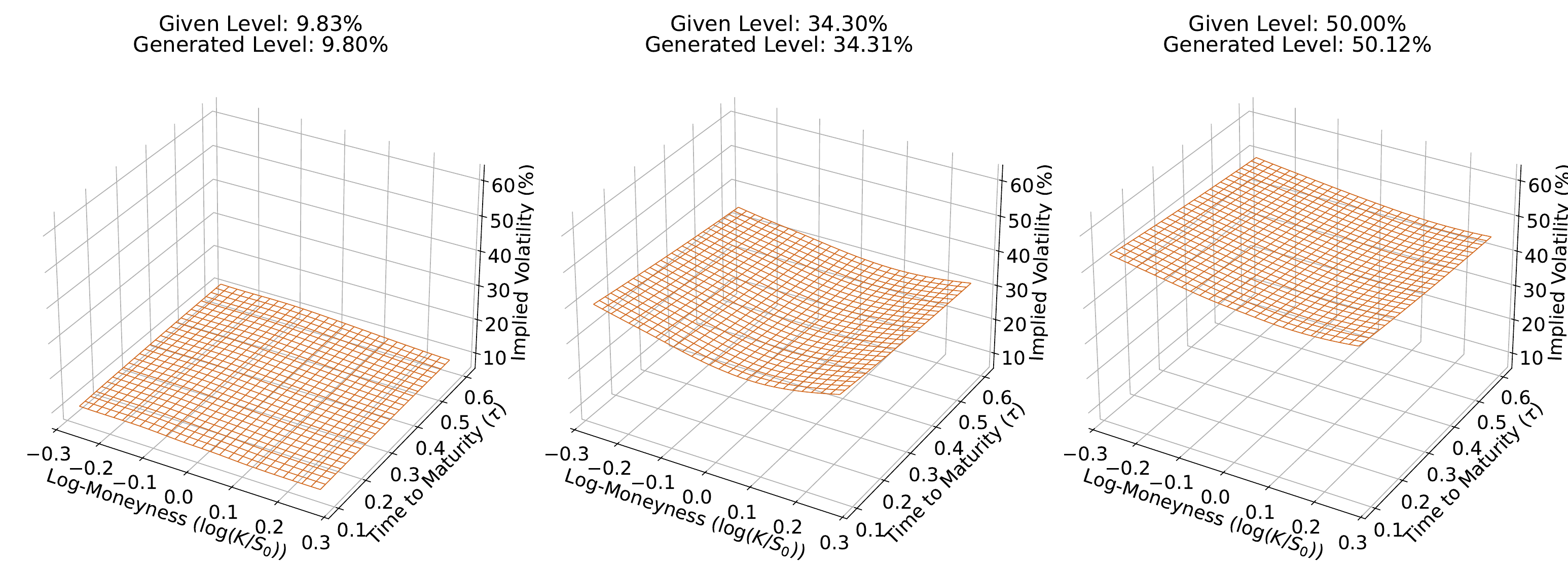}
    \caption{3D example: generated IVS with a pronounced ``smile'' while controlling the \textit{level} feature ($y_L$).}
    \label{fig:3Dsmile}
\end{figure}

\begin{figure}[htbp]
    \centering
    \includegraphics[width=1\linewidth]{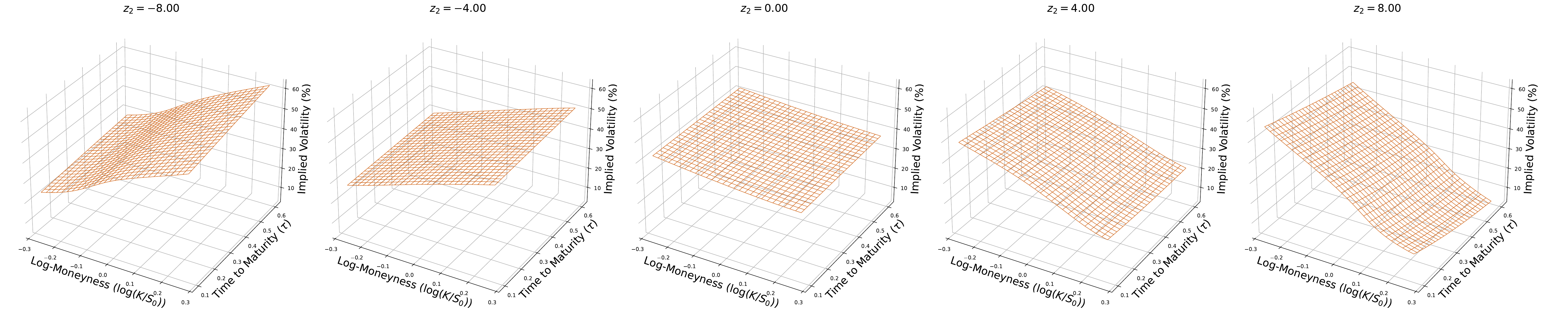}
    \caption{Effect of varying latent dimension $z_2$ on the generated IVS with other latent variables set to zero.}
    \label{fig:z2_ivs_effect}
\end{figure}

\begin{figure}[htbp]
    \centering
    \includegraphics[width=1\linewidth]{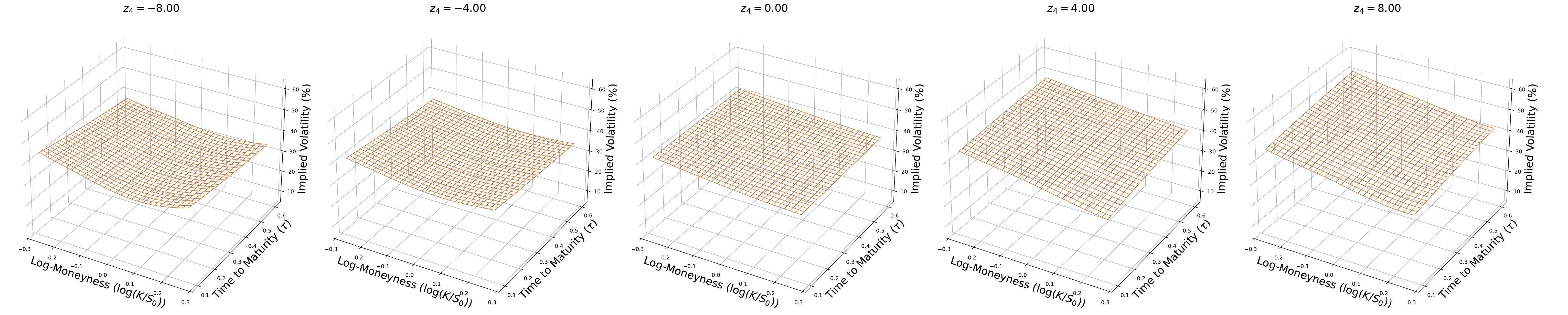}
    \caption{Effect of varying latent dimension $z_4$ on the generated IVS with other latent variables set to zero.}
    \label{fig:z4_ivs_effect}
\end{figure}

\newpage
\subsection{Experiment II} \label{sec:exp2}
\begin{figure}[htbp]
    \centering
    \includegraphics[width=1\linewidth]{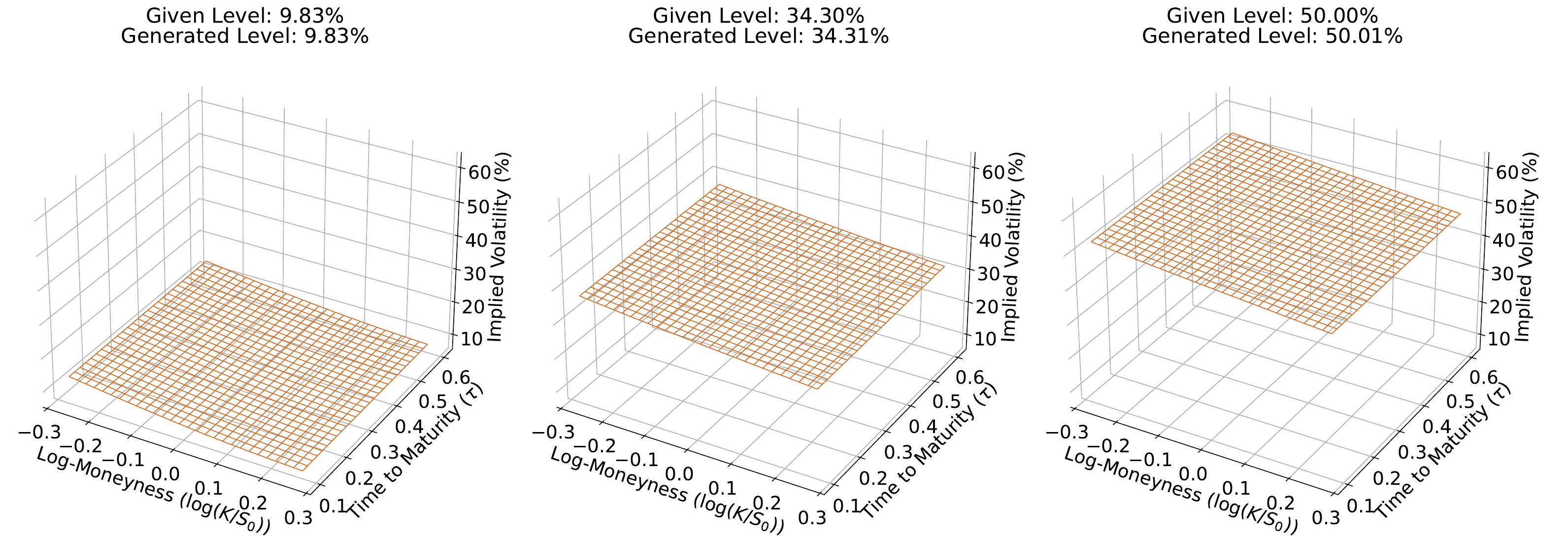}
    \caption{3D visualization of controlled \textit{level} variations under three-feature control with fixed latent variables.}
    \label{fig:3dl}
\end{figure}

\begin{figure}[htbp]
    \centering
    \includegraphics[width=1\linewidth]{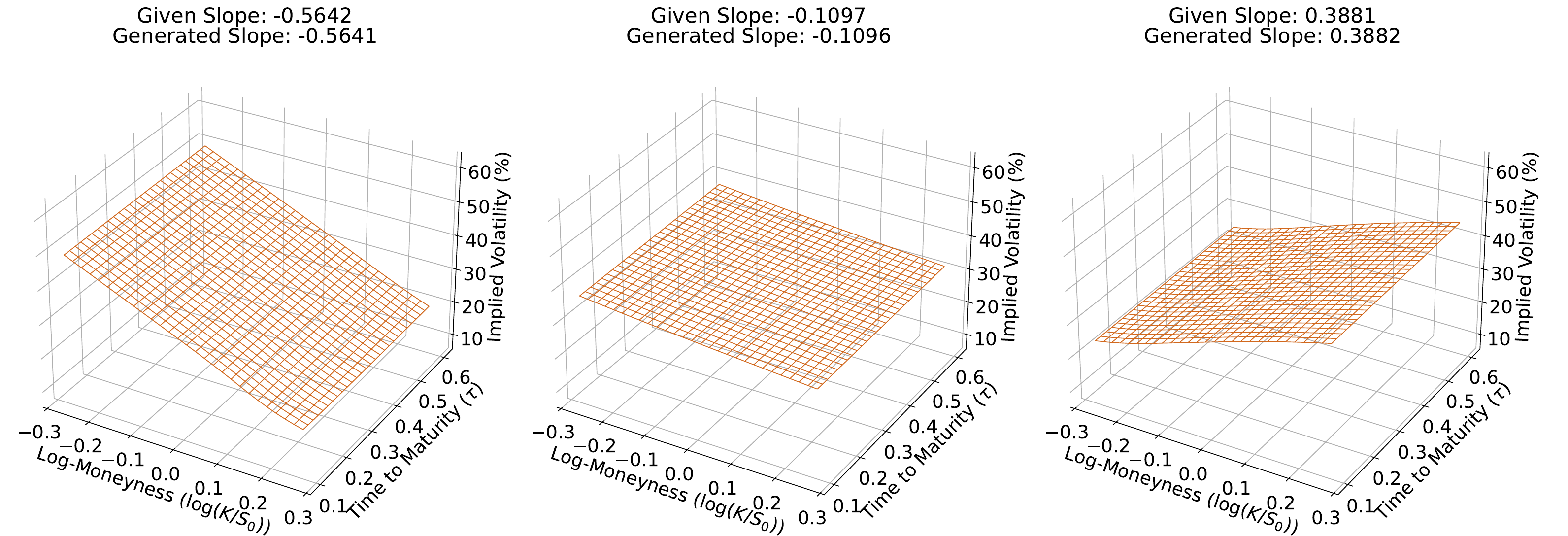}
    \caption{3D visualization of controlled \textit{slope} variations under three-feature control with fixed latent variables.}
    \label{fig:3ds}
\end{figure}

\begin{figure}[htbp]
    \centering
    \includegraphics[width=1\linewidth]{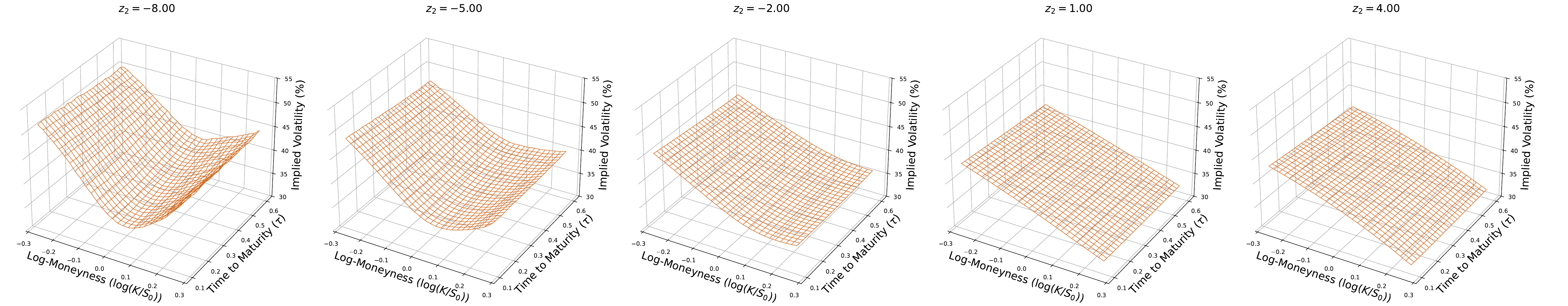}
    \caption{3D visualization of the effect of latent variable $z_2$ under three-feature control ($y_L$, $y_S$, $y_T$) with other latent variables fixed at zero.}
    \label{fig:3dz2LST}
\end{figure}

\begin{figure}[htbp]
    \centering
    \includegraphics[width=\linewidth]{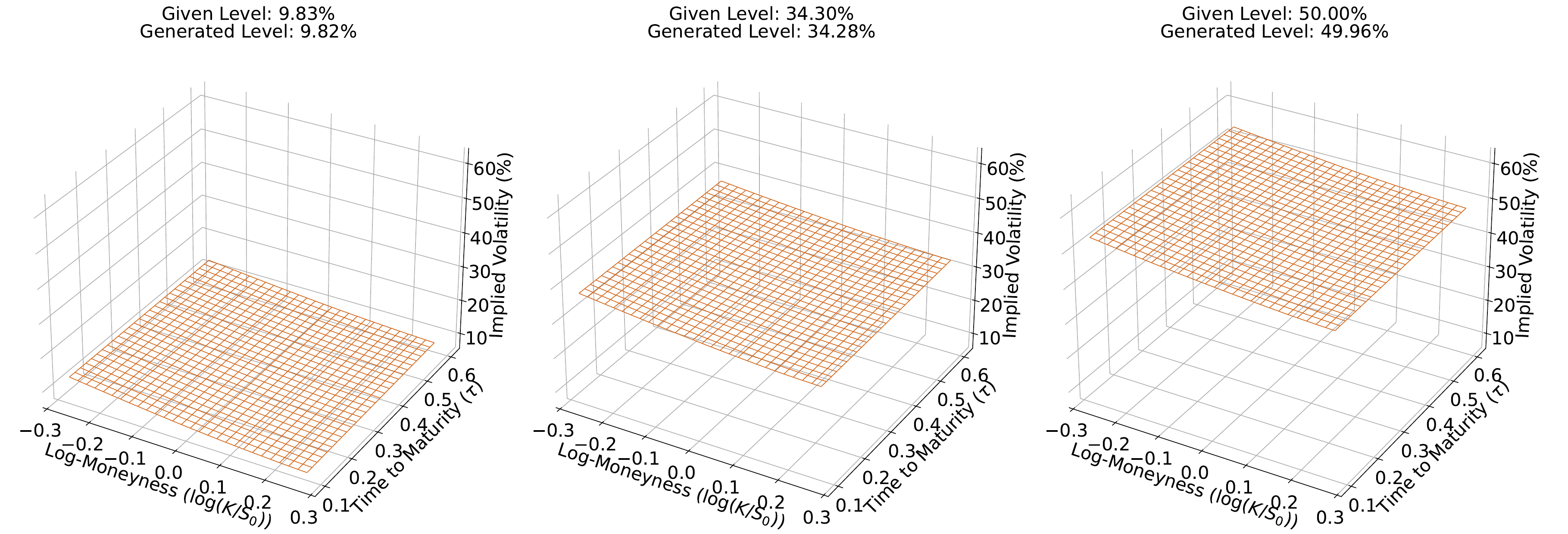}
    \caption{Three-dimensional illustration of controlled variation in the level feature.}
    \label{fig:3d_level_LSCT}
\end{figure}

\begin{figure}[htbp]
    \centering
    \includegraphics[width=\linewidth]{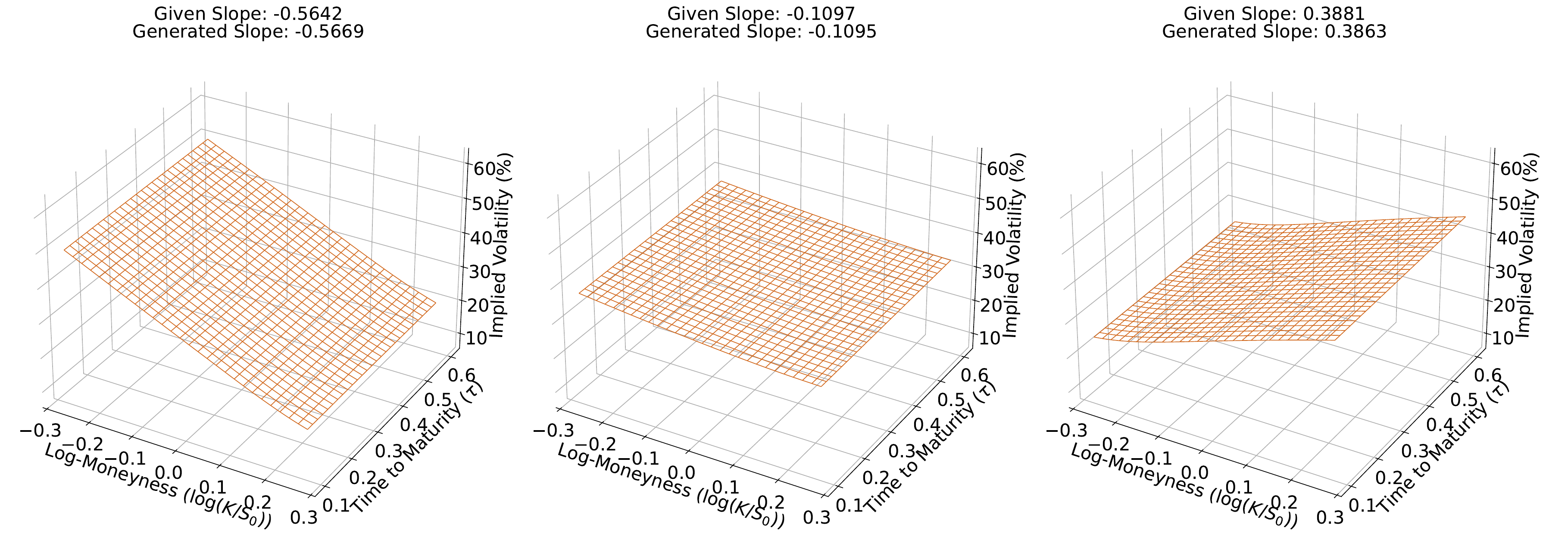}
    \caption{Three-dimensional illustration of controlled variation in the slope feature.}
    \label{fig:3d_slope_LSCT}
\end{figure}

\begin{figure}[htbp]
    \centering
    \includegraphics[width=\linewidth]{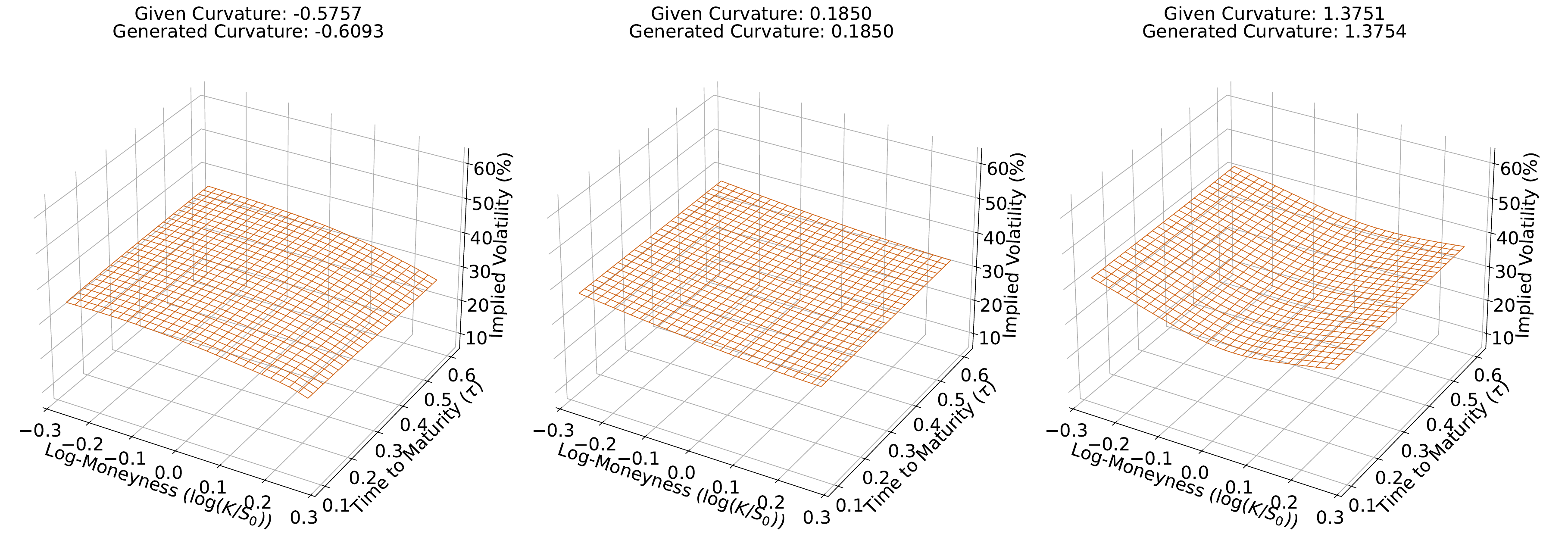}
    \caption{Three-dimensional illustration of controlled variation in the curvature feature.}
    \label{fig:3d_curvature_LSCT}
\end{figure}

\begin{figure}[htbp]
    \centering
    \includegraphics[width=\linewidth]{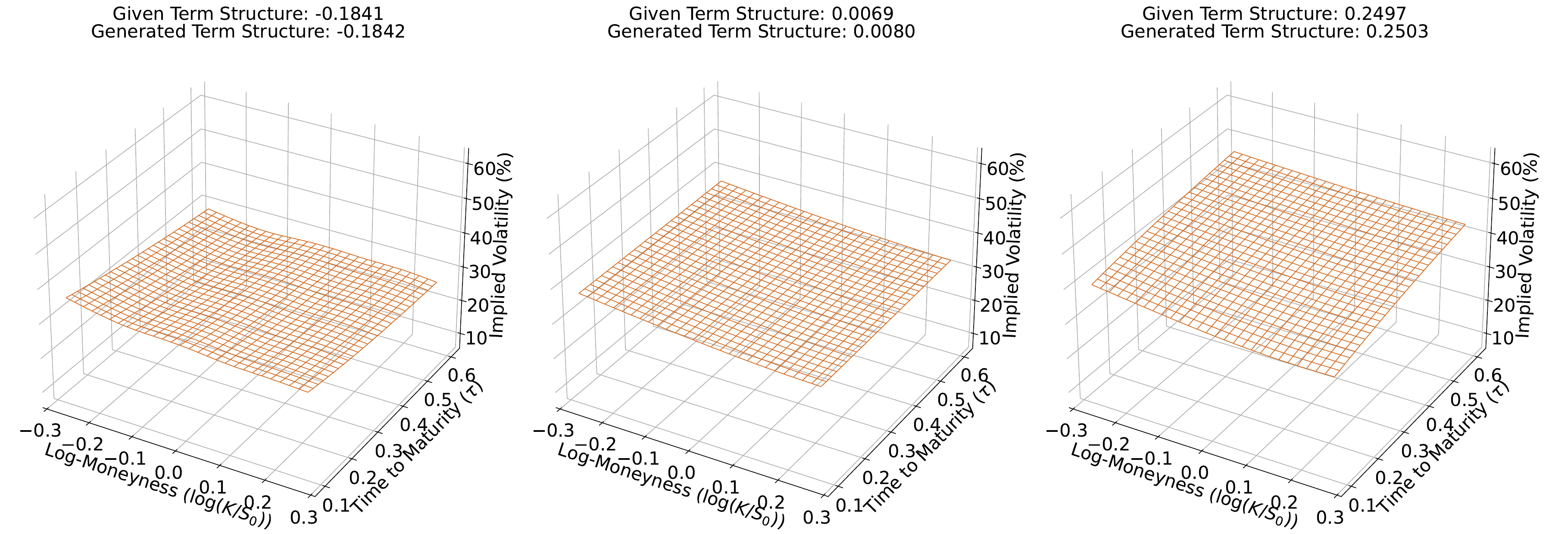}
    \caption{Three-dimensional illustration of controlled variation in the term structure feature.}
    \label{fig:3d_term_structure_LSCT}
\end{figure}

\begin{figure}[htbp]
    \centering
    \includegraphics[width=\linewidth]{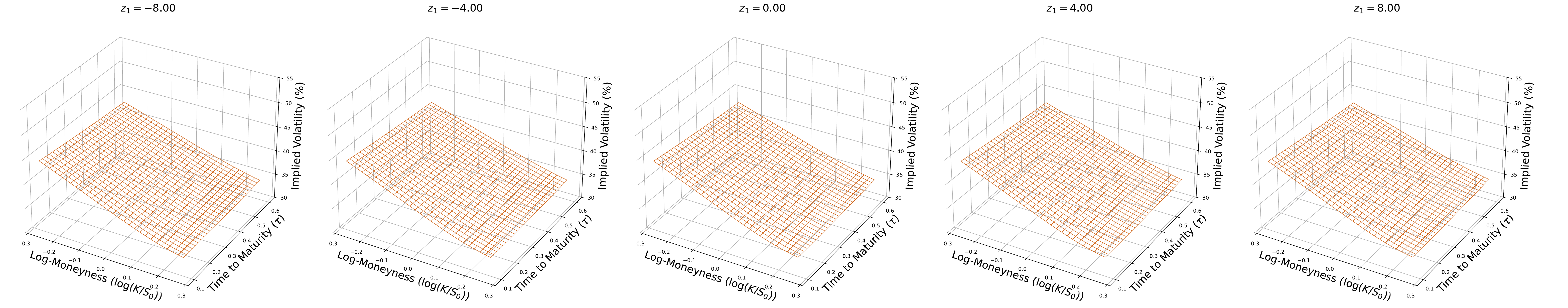}
    \caption{Impact of varying latent variable \( z_1 \) on IVSs under complete four-feature control scenario.}
    \label{fig:z1_effect_LSCT}
\end{figure}

\begin{figure}[htbp]
    \centering
    \includegraphics[width=1\linewidth]{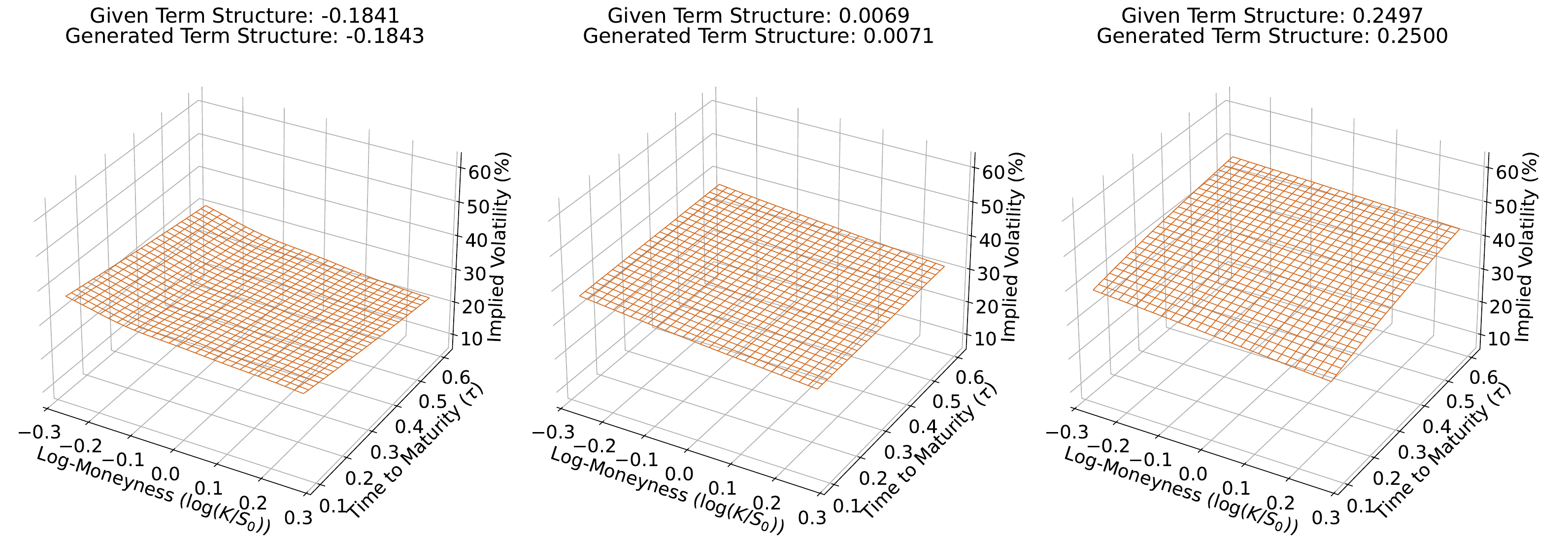}
    \caption{3D visualization of controlled ``term-structure'' variations under three-feature control with fixed latent variables.}
    \label{fig:3dt}
\end{figure}

\end{document}